\documentclass[11pt,a4paper,english]{article}
\usepackage{graphicx}
\usepackage{amssymb,latexsym}
\usepackage{amsmath}
\usepackage{epsf}
\usepackage{pdfsync}
\usepackage{epsfig}
\usepackage[parfill]{parskip}
\usepackage[matrix,arrow,color]{xy}
\usepackage[usenames]{color}
\usepackage{hyperref}
\usepackage{mathrsfs}
\usepackage[font={small}]{caption}
\usepackage{subcaption}
\usepackage[export]{adjustbox}
\usepackage{float}
\usepackage{wrapfig}
\usepackage{microtype}
\usepackage{slashed}

\input{xy}
\xyoption{all}

\input{epsf}

\makeatletter

\usepackage{setspace}

\usepackage{lscape}

\catcode`@=11
\renewcommand{\section}
{\@startsection{section}{1}{0pt}{\medskipamount}{\medskipamount}{\large\bf}}
\makeatletter\renewcommand{\subsection}
{\@startsection{subsection}{2}{\z@}{-3.25ex plus -1ex minus -.2ex}
{1.5ex plus .2ex}{\it }}

\numberwithin{equation}{section}
\catcode`@=12

\newcommand{\ban}{\begin{eqnarray}}
\newcommand{\ean}{\end{eqnarray}}

\def\e{{\,\rm e}\,}

\def\ii{{\,{\rm i}\,}}

\def\beq{\begin{equation}}
\def\bee{\begin{equation}}
\def\eeq{\end{equation}}
\def\bea{\begin{eqnarray}}
\def\eea{\end{eqnarray}}
\def\bd{\begin{displaymath}}
\def\ed{\end{displaymath}}

\newcommand{\Cint}{\int\kern-10.5pt-\kern7pt}

\newcommand{\be}{\begin{equation}}
\newcommand{\ee}{\end{equation}}
\newcommand{\bal}{\begin{align}}
\newcommand{\eal}{\end{align}}

\newcommand\fverbit{\egroup\item[\fbox{\unhbox\pippobox}]}
\newbox\pippobox

{

\def\be{\begin{equation}}
\def\ee{\end{equation}}
\def\bea{\begin{eqnarray}}
\def\eea{\end{eqnarray}}

\begin{document}

\begin{titlepage}
\setcounter{page}{1}

\vskip 5cm

\begin{center}

\vspace*{3cm}

{\Huge On Framed Quivers, BPS Invariants \\[8pt] and Defects}

\vspace{15mm}

{\large\bf Michele Cirafici}
\\[6mm]
\noindent{\em Center for Mathematical Analysis, Geometry and Dynamical Systems,\\
Instituto Superior T\'ecnico, Universidade de Lisboa, \\
Av. Rovisco Pais, 1049-001 Lisboa, Portugal}\\ Email: \ {\tt michelecirafici@gmail.com}

\vspace{15mm}

\begin{abstract}
\noindent

In this note we review some of the uses of framed quivers to study BPS invariants of Donaldson-Thomas type. We will mostly focus on non-compact Calabi-Yau threefolds. In certain cases the study of these invariants can be approached as a generalized instanton problem in a six dimensional cohomological Yang-Mills theory. One can construct a quantum mechanics model based on a certain framed quiver which locally describes the theory around a generalized instanton solution. The problem is then reduced to the study of the moduli spaces of representations of these quivers. Examples include the affine space and noncommutative crepant resolutions of orbifold singularities. In the second part of the survey we introduce the concepts of defects in physics and argue with a few examples that they give rise to a modified Donaldson-Thomas problem. We mostly focus on divisor defects in six dimensional Yang-Mills theory and their relation with the moduli spaces of parabolic sheaves. In certain cases also this problem can be reformulated in terms of framed quivers.
\end{abstract}

\vspace{15mm}


\end{center}
\end{titlepage}


\tableofcontents

\section{Introduction}

In this survey we will discuss various connections between several topics in mathematical physics. The underlying theme is the structure of BPS states on a local Calabi-Yau threefold. The BPS sector of supersymmetric field and string theories consists in quantities which are protected from quantum corrections and can sometime be studied exactly. Objects of this type are extremely important in physics, as they often provide a window into the non-perturbative aspects of these theories, which are usually out of reach with conventional techniques. On the other hand BPS quantities have a mathematical description, often directly in terms of geometrical or algebraic quantities. The interplay between these two perspective has offered beautiful insights in mathematics and in physics.

Donaldson-Thomas theory \cite{thomas,Kontsevich:2008fj} is, broadly speaking, concerned with BPS states in field or string theories with $\mathcal{N}=2$ supersymmetry in four dimensions. Within this class we have, for example, the four dimensional supersymmetric field theories of class $\mathcal{S}$, which arise from compactification of the six dimensional $\mathcal{N}=(2,0)$ theory, or string compactifications on compact or non-compact Calabi-Yau threefolds. We will mostly focus on string compactifications on non-compact Calabi-Yau and use techniques rooted in topological quantum field theory to set up the relevant enumerative problems. The latter have the form of certain quantities, roughly speaking the ``volumes'' of specific moduli spaces which arise in the physical problems. We will generically call these quantities Donaldson-Thomas invariants, or BPS invariants of Donaldson-Thomas type, without any pretense of being exhaustive (especially ignoring altogether the motivic roots of the problem). General aspects of Donaldson-Thomas theory are reviewed in Section \ref{BPS}.

Among the deepest traits of Donaldson-Thomas theory is the wall-crossing behavior of the enumerative invariants. It originates from the property of BPS states to decay or form bound states whose constituents are themselves BPS. The loci where this happens are called walls of marginal stability. This property conjecturally determines an algebraic structure on the space of BPS states. What this algebra precisely is supposed to be is actively debated; a strong candidate is the cohomological Hall algebra of \cite{Kontsevich:2010px}. The wall-crossing property of the BPS invariants is captured by the wall-crossing formulae \cite{Kontsevich:2008fj,Joyce:2008pc,Manschot:2010qz}. In the recent years the uses of the wall-crossing formulae and more in general the associated structure have vastly enhanced our understanding of the BPS sector of field and string theories. 

The relevant moduli spaces of vacua are therefore divided into chambers by the walls of marginal stability. Each chamber contains an usually challenging enumerative problem associated with the stable BPS invariants. A general, perhaps naive, strategy to attack the problem is to start from  ``easier'' chambers and then move along using the wall-crossing formulae. In this survey we will mainly discuss two of these chambers, within the context of non-compact Calabi-Yaus. 

One of these chambers will be at ``large radius'', where string theory corrections to ordinary algebraic geometry are negligible. The BPS states we will consider are parametrized by torsion free or ideal sheaves. Thus the problem reduces to construct an appropriate moduli space of sheaves and define appropriate BPS invariants. These issues will be discussed in Sections \ref{cohomological}-\ref{localization}. The second chamber will be the ``noncommutative resolution chamber'', where the local threefold develops a singularity and the manifold structure breaks down. Nevertheless physics can be formulated in terms of certain noncommutative algebras and the relevant moduli problems are problems in representation theory. We will introduce and develop these concepts in Sections \ref{noncommutative}-\ref{instantons}. In both of these chambers one can make some progress by using field and string theory concepts. In particular all of these problems can be approached from the point of view of a certain six dimensional topological Yang-Mills theory \cite{Iqbal:2003ds,Cirafici:2008sn}, where the relevant moduli spaces arise as moduli spaces of generalized instanton configurations. In this survey we will highlight those concepts which originate from quivers and their representations, in accordance with the theme of this volume. In both chambers the relevant BPS invariants can be understood from the point of view of a supersymmetric quiver quantum mechanics which localizes onto the moduli spaces of BPS configurations. This quiver quantum mechanics is a physical counterpart of studying the intersection theory of the generalized instanton moduli spaces via equivariant localization \cite{Cirafici:2008sn,Cirafici:2008ee,Cirafici:2010bd,Cirafici:2011cd,Cirafici:2012qc}.

In the second part of this survey, Sections \ref{divisor}-\ref{line}, we will introduce and study defects. A defect can be understood as imposing certain boundary conditions on the physical fields, for example along a line or a surface. In this case we talk about a line or surface defect. It is natural to wonder if given a certain moduli problem associated with a physical theory and its BPS invariants, one can introduce new enumerative invariants by introducing defects \cite{Gaiotto:2010be,Gaiotto:2011tf,Chuang:2013wt,Cirafici:2013nja}. The idea is that the presence of a defect modifies the relevant moduli spaces by imposing certain conditions, or restricting one's attention only to certain configurations. We will do so in certain particular cases and argue that indeed one can try to define Donaldson-Thomas type of invariants out of these moduli spaces. Our prime example will be the case of divisor defects in a six dimensional field theory \cite{Cirafici:2013nja}. We will argue that such a defect requires the physical configurations to correspond to parabolic sheaves. In particular we will see that in a particularly simple case, the conjectured new invariants can be studied in great detail, via certain classes of quivers. We will furthermore discuss how these ideas also apply to other cases, such as in higher dimensions or in the context of line defects in four dimensional field theories \cite{Cirafici:2013bha,CDZ}.

In this survey we take an expository tone, often referring the reader to the original literature for explicit details and focusing more on the general concepts and ideas. In particular we will stress the role played by quivers and their representations, often neglecting other (important) issues. The purpose of this survey is not to present the material in a self-contained manner, but rather provide an entry point to the (by now rather vast) literature.

\section{BPS states and Donaldson-Thomas theory} \label{BPS}

We begin with a brief discussion of BPS states on Calabi-Yau threefolds and its relation with Donaldson-Thomas theory. Fix a threefold $X$ and consider the type IIA string compactified over $X$. The effective theory in four dimensions has $\mathcal{N}=2$ supersymmetry. A BPS state preserves half of these supersymmetries. Their descriptions depends sensitively on the moduli of the Calabi-Yau $X$. We start by considering the large radius approximation. The BPS states are labelled by a charge vector $\gamma \in \Gamma_g$ and in this case the charge lattice is given by the cohomology groups of the threefold
\begin{equation}
\gamma \ \in \ \Gamma_g = \Gamma_g^{\rm m} \oplus \Gamma_g^{\rm e} = \left( H^0 (X ,
  \mathbb{Z}) \oplus H^2 (X , \mathbb{Z}) \right) \ \oplus \ \left( H^4 (X , \mathbb{Z})
  \oplus H^6 (X , \mathbb{Z}) \right) \ ,
\end{equation}
which can be separated into the electric and magnetic charge sublattices
$\Gamma_g^{\rm e}$ and $\Gamma_g^{\rm m}$. Physically these lattices correspond to the charges of $D$-branes wrapping $p$-cycles of $X$
\begin{equation}
{\rm D}p \ \longleftrightarrow \ H^{6-p} (X , \mathbb{Z}) = H_{p} (X , \mathbb{Z}) \ , \qquad p = 0,2,4,6 \ ,
\end{equation}
where we have used Poincar\'e duality (when the Calabi-Yau $X$ is non-compact, as will be the case in most of this survey, this discussion can be rephrased in terms of cohomology with compact support). 
In this class of compactification the central charge of the supersymmetry algebra has an explicit form dictated by the geometry of the threefold. At large radius
\begin{equation}
Z_X (\gamma;t) = -\int_{X}\, \gamma \wedge \e^{-t}
\end{equation}
gives the central charge of the state $\gamma$. Here $t = B+ \ii J$ is the complexified K\"ahler modulus consisting of the background supergravity Kalb-Ramond two-form $B$-field and the K\"ahler $(1,1)$-form $J$ of~$X$.

A measure of the degeneracy of BPS states is given by the Witten index
\begin{equation} \label{Windex}
\Omega_X \left( \gamma  \right) = \mathrm{Tr}_{\mathcal{H}^X_{\gamma , \mathrm{BPS}}}\, (-1)^F
\end{equation}
given in terms of a trace over the fixed charge sectors of the single-particle Hilbert space, defined as
\begin{equation}
\mathcal{H}_{\mathrm{BPS}}^X = \bigoplus_{\gamma \in \Gamma} \, \mathcal{H}^X_{\gamma ,
  \mathrm{BPS}} \ .
\end{equation}
$F$ is a certain operator acting on one-particle states with charge $\gamma$, which originates from the isometry group of the four dimensional effective supersymmetric theory.

Note that the definition of a BPS state as well as the definition of the Witten index are rather straightforward and purely based on representation theory arguments. The BPS Hilbert spaces are finite dimensional and decompose according to the representation theory of the symmetry group of the four dimensional effective $\mathcal{N}=2$ theory and, somewhat oversimplifying, the Witten index just counts with signs the multiplicities of these representations. On the other hand thanks to the string compactifications these quantities are related to geometrical structures within the Calabi-Yau $X$. This correspondence is at the core of many beautiful connections between mathematics and physics. Roughly speaking Donaldson-Thomas theory is the mathematical counterpart of these physical facts.

We can consider for example a particular situation where $X$ is a non-compact threefold and the charge vector is $\gamma = (1 , 0 , - \beta , n)$. Physically these configurations corresponds to bound states of a gas of $D0$ and $D2$ branes with a single $D6$ brane wrapping the whole threefold. Mathematically the relevant geometrical configurations on $X$ are ideal sheaves.

An ideal sheaf $\mathcal{I}$  is a torsion free sheaf of rank one with trivial determinant. Since the determinant is trivial, the double dual $\mathcal{I}^{\vee \vee}$ is isomorphic to the trivial bundle and in particular $c_1 (\mathcal{I})= 0$. The condition that the sheaf is torsion free means that it can be embedded in a bundle; roughly speaking an ideal sheaf can be thought of as an object which fails to be a line bundle only on a finite set of singularities. There is a correspondence between ideal sheaves and schemes given by the short exact sequence
\begin{equation} \label{idealseq}
\xymatrix@1{
0 \ \ar[r] &  \ \mathcal{I} \ \ar[r] & \ \mathcal{O}_X \ \ar[r] & \ \mathcal{O}_Y \ \ar[r] & \
0 \ .
}
\end{equation}
Here $Y$ is a subscheme of $X$ and the sequence implies that we can think of an ideal sheaf as the kernel of the restriction map $\mathcal{O}_X \rightarrow \mathcal{O}_Y$ of structure sheaves.
To define Donaldson-Thomas invariants, we look at $\mathcal{M}^{\rm BPS}_{n,\beta} (X)$ the moduli space of ideal sheaves such that
\begin{equation}
\chi(\mathcal{I})=n \qquad \mbox{and} \qquad \mathrm{ch}_2(\mathcal{I})=-\beta \ .
\end{equation}
Due to (\ref{idealseq}) we can identify this moduli space with the  Hilbert scheme $\mathrm{Hilb}_{n,\beta} (X)$ of points and curves on $X$, which parametrizes subschemes $Y\subset X$ with no component of codimension
one and such that
\begin{eqnarray}
n=\chi (\mathcal{O}_Y) \qquad \mbox{and} \qquad
\beta = [Y] \ \in \ H_2 (X ,\mathbb{Z}) \ ,
\end{eqnarray}
where $\chi$ denotes the holomorphic Euler characteristic. Donaldson-Thomas invariants are defined via integration over these moduli spaces as
\begin{equation} \label{DTdef}
\Omega_X(n,\beta)={\tt DT}_{n,\beta}(X) := \int_{[\mathcal{M}^{\rm BPS}_{n,\beta} (X)]^{\mathrm{vir}}} \, 1 \ .
\end{equation}
To properly define integration over this, and other, moduli schemes, one needs to define a virtual fundamental cycle. Roughly speaking while these moduli spaces are in general not manifolds, they behave as if they were, at least generically. In these cases one can define a virtual fundamental class, which depends on the deformation and obstruction theory of the moduli spaces \cite{fantechi} . In our cases of interest the deformation complex of the moduli space contains two terms, parametrizing obstructions and deformations, and plays the role of a cotangent complex to the moduli space, therefore providing a notion of integration (much as it happens for the ordinary cotangent space to a smooth manifold, out of which the integration measure is constructed). Furthermore in our cases the obstruction space and the deformation space are dual to each other in a suitable sense, and the virtual fundamental class has degree zero. This means that integrating $1$ provides a virtual counting of points in $\mathcal{M}^{\rm BPS}_{n,\beta} (X)$. A more thorough discussion can be found in~\cite{thomas,fantechi} (and~\cite{Szabo:2009vw,Szabo:2011mj} within the present context). 

An alternative formulation is due to Behrend~\cite{behrend} and regards Donaldson-Thomas invariants as the \emph{weighted} topological Euler characteristics
\begin{equation}
{\tt DT}_{n,\beta}(X) = \chi \big( \mathcal{M}^{\rm BPS}_{n,\beta} (X) \, , \, \nu_X \big) = \sum_{n\in\mathbb{Z}}\, n \
\chi\big(\nu_X^{-1}(n)\big) \ ,
\label{Behrend}\end{equation}
where $\nu_X:\mathcal{M}^{\rm BPS}_{n,\beta} (X) \to\mathbb{Z}$ is a canonical constructible function.

This enumerative problem is very rich and interesting. However it is only the tip of the iceberg, as it is only one of the many enumerative problems predicted by physics. As the physical parameters are varied, some physical states may become unstable and decay, or form stable bound states with other states. When this happens the Hilbert space $\mathcal{H}_\gamma^X$ over which the Witten index is defined, will gain or lose a factor. As a consequence the degeneracies of BPS states will jump, a phenomenon known as wall-crossing. Such jumps occur at walls of marginal stability, loci where $\mathrm{Arg} (Z_X (\gamma_1 , t)) = \mathrm{Arg} (Z_X (\gamma_2 , t))$, and the jump in the indices is governed by a wall crossing formula.

As a consequence the K\"ahler moduli space of the threefold $X$ is divided into chambers, each one with its BPS spectrum of states, and therefore each one associated with an enumerative problem. The full mathematical theory of Donaldson-Thomas invariants need rather sophisticated tools \cite{Kontsevich:2008fj,Kontsevich:2010px}, while in this survey we will limit ourselves to give a simplified treatment. Roughly speaking BPS states on a Calabi-Yau threefold are properly described in terms of the derived category of coherent sheaves $\mathcal{D}^b (X)$, with stability conditions given by an ordering of the central charge phases of the BPS states. In many cases however a description in terms of an abelian subcategory $\mathcal{A}$ is sufficient.

In this survey we will focus only on two chambers where simpler tools can be used, grounded in quantum field theory. The reason is that in these two chambers one can make very concrete computational progress. The chambers in question are that called ``at large radius'' and the ``noncommutative crepant resolution chamber''. In the first one the relevant abelian subcategory is the category of coherent sheaves and one can make use of  geometrical concepts such as sheaves or bundles defined on a smooth Calabi-Yau. The second chamber corresponds to the situation where the Calabi-Yau develops a singularity, for example an orbifold or a conifold singularity; the relevant abelian subcategory is the category of representations of a certain quiver, and one has at disposal many tools based on algebra or representation theory.

\section{Cohomological gauge theory} \label{cohomological}

At large radius the problem of studying Donaldson-Thomas invariants on a Calabi-Yau manifold $X$ can be approached via a cohomological gauge theory \cite{Iqbal:2003ds,Cirafici:2008sn}. This is a particular version of a topological quantum field theory obtained by the topological twist of six dimensional Yang-Mills theory. We can introduce this theory as follows: the bosonic sector consists of a connection $A$ on a $G$-bundle $\mathcal{E} \longrightarrow X$, the complex one form Higgs field $\Phi$ taking values in the adjoint bundle $\mathrm{ad} \, \mathcal{E}$, and the 3-form fields $\rho^{(3,0)}$ and $\rho^{(0,3)}$. The fermionic sector is twisted, by which we mean that the fermions can be identified with differential forms thanks to the isomorphism between the spin bundle and the bundle of differential forms $\mathcal{S} (X) \simeq \Omega^{0,\bullet} (X)$, which in particular holds for Calabi-Yau manifolds. The overall fermionic content is $(\eta , \psi^{(1,0)} , \psi^{(0,1)},\chi^{(2,0)}, \chi^{(0,2)}, \psi^{(3,0)},\psi^{(0,3)})$, where $\eta$ is a complex scalar and we have written down explicitly the form degree. The bosonic part of the action is 
\begin{align}
S = \, & \frac12 \int_X \mathrm{Tr} \left(
\mathrm{d}_A \Phi \wedge * d_A \overline{\Phi} + \left[ \Phi , \overline{\Phi} \right]^2 + |F_A^{(0,2)} + \overline{\partial}^\dagger_A \rho |^2 + |F^{(1,1)}_A|^2
\right)
\\ \nonumber
&  + \frac12 \frac{1}{(2 \pi)^2} \int_X \mathrm{Tr} \left( F_A \wedge F_A \wedge t + \frac{\lambda}{6 \pi} F_A \wedge F_A \wedge F_A \right)
\end{align}
Here $F_A = \mathrm{d} \, A + A \wedge A$ is the gauge field strength, $\mathrm{d}_A = \mathrm{d} + A$ the covariant derivative, and the Hodge star operator $*$ is taken with respect to the K\"ahler metric of $X$. The parameter $\lambda$ is a coupling constant which from a string theory perspective should be thought of as the topological string coupling. 

The gauge theory localizes onto the moduli space of solutions of the ``generalized instanton'' equations
\begin{eqnarray} \label{inste} F_A^{(0,2)} &=&
  \overline{\partial}\,_A^{\dagger} \rho
  \ , \nonumber\\[4pt]
F_A^{(1,1)} \wedge t \wedge t + \big[\rho\,,\, \overline{\rho}\,\big] &=&
l~t \wedge t \wedge t \ , \nonumber\\[4pt] \mathrm{d}_A \Phi &=& 0 \ .
\end{eqnarray}
On a Calabi-Yau variety we can set $\rho=0$ without loss of generality. In this case the first two equations of (\ref{inste}) become the Donaldson-Uhlenbeck-Yau equations which parametrize holomorphic vector bundles $\mathcal{E}$ on $X$. The parameter $l$ is proportional to the degree of the bundle $\mathcal{E}$; unless stated otherwise we will for simplicity set it to zero. 

We want to use gauge theory techniques to study the moduli space of holomorphic bundles, following the approach of \cite{Nekrasov:2002qd}. However to obtain a better behaved moduli space it is customary to enlarge the set of allowed configurations to include torsion free sheaves. In the following we will denote by $\mathcal{M}^{\mathrm{inst}}_{n,\beta;r} $ the moduli space of torsion free coherent sheaves $\mathcal{E}$ with characteristic classes $(\mathrm{ch}_3 (\mathcal{E}) , \mathrm{ch}_2 (\mathcal{E}))=(n,-\beta)$ and rank $r$. To understand the local geometry of these moduli spaces, we consider the instanton deformation complex
\begin{equation} \label{defcomplex}
\xymatrix@=6mm{0 \ar[r] & \Omega^{0,0} ( X , \mathrm{ad}\, \mathcal{E})
\ar[r]^{\hspace{-1.3cm} C} & ~\Omega^{0,1} ( X , \mathrm{ad}\, \mathcal{E}
) \oplus \Omega^{0,3} ( X , \mathrm{ad}\, \mathcal{E})
\ar[r]^{\hspace{1.3cm} D_A} & \Omega^{0,2} ( X , \mathrm{ad}\, \mathcal{E}) \ar[r] & 0 } \ .
\end{equation}
Here $C$ represents a linearized complexified gauge transformation, while $D_A$ the linearization of the first equation of (\ref{inste}). The cohomology of the complex in degree zero corresponds to reducible connections, and we will assume it vanishes. The cohomology in degree one is the Zarinski tangent space to $\mathcal{M}^{\mathrm{inst}}_{n,\beta;r} $ at a point corresponding to a sheaf $\mathcal{E}$, and the cohomology at degree two represent the obstruction bundle (or normal bundle) $\mathcal{N}_{n,\beta;r}$. The gauge theory partition function can be written as a sum over topological sectors; each sector contributes with an integral over the moduli space of holomorphic bundles where the integration measure is given by the Euler class of the obstruction bundle $\mathrm{eul} (\mathcal{N}_{n,\beta;r})$. To write this partition function, we pick a basis of $H_2 (X , \mathbb{Z})$ and expand the class $\beta = \sum_i \, n_i \, S_i$, where $i=1, \dots, b_2 (X)$. Then we set $Q_i = \mathrm{e}^{-t_i}$ with $t_i = \int_{S_i} \, t$ and define $Q^{\beta} := \prod_i \, Q_i^{n_i}$. Therefore we can write
\begin{equation} \label{Zdt}
Z^X_\mathrm{gauge} (q , Q ; r) = \sum_{k \, \beta} \ q^k \, Q^{\beta} \ \int_{\mathcal{M}^{\mathrm{inst}}_{n,\beta;r} } \ \mathrm{eul} (\mathcal{N}_{n,\beta;r})
\end{equation}
Due to the expository tone of this survey, we will refrain to properly define the integrals over the moduli spaces, except in special cases. These integrals represents Donaldson-Thomas invariants; although they can be defined in full generality, we will see that computational progress can be made only in special occasions, such as toric varieties and $U(1)^r$ gauge theories, where equivariant localization formulae can be applied. 

\section{Quivers} \label{quivers}

Several aspects of Donaldson-Thomas theory on local threefolds can be understood from an algebraic perspective using quivers \cite{szendroi,Cirafici:2010bd,Cirafici:2011cd}. A quiver is a finite directed graph, consisiting in a quadrupole $(\mathsf{Q}_0 , \mathsf{Q}_1 , t,s)$; here $\mathsf{Q}_0$ and $\mathsf{Q}_1$ are two finite sets, representing the nodes and the arrows respectively, while the maps $s,t \, : \, \mathsf{Q}_1 \longrightarrow \mathsf{Q}_0$ associate to each arrow $a \in \mathsf{Q}_1$ its starting vertex $s (a) \in \mathsf{Q}_0$ or its terminal vertex $t (a) \in \mathsf{Q}_0$. To the set of arrows one can associate a set of relations $\mathsf{R}$. To a quiver we can associate its path algebra $\mathsf{A} = \mathbb{C} \mathsf{Q} / \langle \mathsf{R} \rangle$, defined as the algebra of paths modulo the ideal generated by the relations. A path is defined as a set of arrows which compose; the product in the algebra is the concatenation of paths where possible, or zero otherwise. A relation in the path algebra is a $\mathbb{C}$-linear combination of paths. 

In most physical applications, the relations $\mathsf{R}$ are derived from a superpotential $\mathsf{W}$. This is a function $\mathsf{W} \, : \, \mathsf{Q}_1 \longrightarrow \mathbb{C} \mathsf{Q}$ given by a sum of cyclic monomials. We define a differential $\partial_a$ respect to the arrow $a \in \mathsf{Q}_1$ by cyclically permuting the elements of each monomial until the arrow $a$ is in the first position, and then deleting it; differentiation by  $\partial_a$ gives zero if the arrow $a$ is not part of a monomial. If the quiver is equipped with a superpotential, the ideal of relations is given by $\mathsf{R} = \langle \partial_a \mathsf{W} \, \vert \, a \in \mathsf{Q}_1 \rangle$.

A representation of a quiver is defined by the assignement of a complex vector space to each node and a collection of maps between the vector spaces associated with the set of arrows, compatible with the relations $\mathsf{R}$. More precisely to each node $i \in \mathsf{Q}_0$ we associate the vector space $V_i$ of dimension $\mathrm{dim} \, V_i = n_i$, and to each arrow $a \in \mathsf{Q}_1$ a morphism $B_a \in \mathrm{Hom} (V_{s(a)} , V_{t(a)})$, compatible with $\mathsf{R}$. We will denote by $\mathsf{Rep} (\mathsf{Q} , \mathsf{R})$ the category of representations of the quiver $\mathsf{Q}$ with relations $\mathsf{R}$; it can be shown that this category is equivalent to the category $\mathsf{A}-\mathsf{mod}$ of left $\mathsf{A}$-modules. In most physical applications, the objects of interest are actually isomorphism class of representations, which are defined as orbits with respect to the action of the gauge group $\prod_{i \in \mathsf{Q}_0} \, GL (V_i , \mathbb{C})$.

To obtain better behaved moduli spaces, one can modify this construction by framing the quiver. A way of doing so consists in adding to the quiver $\mathsf{Q}$ an extra vertex $\{ \bullet \}$ together with an additional arrow $a_{\bullet}$ such that $s (a_{\bullet}) = \bullet$ and $t (a_{\bullet}) = i_0$, where $i_0 \in \mathsf{Q}_0$ is a reference node of $\mathsf{Q}$. This procedure gives a new quiver $\widehat{\mathsf{Q}}$, defined by $\widehat{\mathsf{Q}}_0 = \mathsf{Q}_0 \cup \{ \bullet \}$ and $\widehat{\mathsf{Q}}_1 = \mathsf{Q}_1 \cup \{ \bullet \xrightarrow{a_\bullet} i_0 \}$. Similarly we can define the path algebra $\hat{A}$ of the framed quiver and framed quiver representations. The notion of framing generalizes immediately to more framing nodes. We will denote framed quivers by $\widehat{\mathsf{Q}}$, no matter the number of framed nodes. 

\section{Donaldson-Thomas invariants, Quiver Quantum Mechanics and Localization} \label{donaldson}

We will begin the discussion of (\ref{Zdt}) in the simplest possible case: when $X = \mathbb{C}^3$ and the gauge symmetry is broken down to its maximal torus $U(1)^r$. In this case we can give an explicit definition of the integrals which appear in (\ref{Zdt}), following what we have explained in Section \ref{BPS}. Furthermore this case allows for explicit computations using techniques of equivariant localization. Indeed there is a natural toric action on $\mathbb{C}^3$ which can be used to localize the integrals in (\ref{Zdt}) onto a finite set of fixed points. In this Section we will discuss the essential points of this procedure.

The main idea is to study the theory around a BPS configuration. This is a standard procedure in physics and mathematics and consists in the construction of an appropriate parametrization of the relevant moduli space of solutions of (\ref{inste}) with fixed characteristic classes. In this case one can construct an explicit parametrization of the moduli space via a generalization of the ADHM construction \cite{Cirafici:2008sn}. This consists of a collection of matrices obeying a set of generalized ADHM equations. Therefore the local parametrization of the moduli space has the form of a matrix model; since this parametrization is explicit this matrix model can be used to compute geometric quantities within the moduli space. It turns out that this matrix model has a very specific form and is given in terms of a topological quiver quantum mechanics. This means that the collection of fields and equations which parametrize the moduli space can be encoded in a representation of a quiver, which we have introduced in Section \ref{quivers}. 

Based on a set of generalized ADHM equations, one can construct a topological quiver quantum mechanics which provides a concrete tool to compute the integrals in (\ref{Zdt}). The homological data of the generalized ADHM construction are encoded in the framed quiver
\begin{equation} \label{ADHMquiver}
\xymatrix@C=20mm{
& \ V \ \bullet \ \ar@(ul,dl)_{B_2} \ar@(ur,ul)_{B_1} \ar@(dr,dl)^{B_3}
\ar@{.>}@(ur,dr)^< < < <{\varphi} & \ \bullet \ W \ar@//[l]_{ \ \ \ \ \ I} 
} \ .
\end{equation}
We will set $\mathrm{dim}_{\mathbb{C}} \, V = n$ and $\mathrm{dim}_{\mathbb{C}} \, W = r$. It is sometimes useful to keep in mind a string theory perspective, where this quiver describes a possible bound state of $n$ $D0$ branes with $r$ $D6$ branes; equivalently from the point of view of the effective topological action on the $D6$ brane worldvolume, $r$ is the rank of the gauge theory and $n$ the instanton number of a gauge field configuration. In (\ref{ADHMquiver}) we have introduced the maps
\begin{eqnarray}
(B_1 , B_2 , B_3 , \varphi) \in \mathrm{Hom}_\mathbb{C} (V,V) \qquad \mbox{and} \qquad
I\in \mathrm{Hom}_\mathbb{C} (W,V) \ .
\label{bosnaive}\end{eqnarray}
We will be interested in configurations of maps where $\varphi$ is trivial; in the quiver quantum mechanics the field $\varphi$ corresponds to degrees of freedom which originate from the six dimensional field $\rho^{(3,0)}$ in (\ref{inste}). The fields in (\ref{bosnaive}) have natural transformations under $U(n)$ and $U(r)$. 

Since the topological quiver quantum mechanics is built out of a parametrization of the instanton moduli space any geometrical quantity within the moduli space is realized as an observable. In particular the partition function of the quantum mechanics computes the volumes of the moduli spaces given by the integrals in (\ref{Zdt}). We will now see this in some detail. 

Since the quiver quantum mechanics is topological, it localizes onto the fixed loci of the BRST charge $\mathcal{Q}$. Therefore the computation of the partition function amounts in classifying the fixed loci of the BRST charge $\mathcal{Q}$ and then computing the contribution around each locus, as a ratio of functional determinants. This approach was discussed in generality in \cite{Moore:1997dj,Moore:1998et}. However we will follow a slightly different route: we modify the BRST charge $\mathcal{Q}$ in an appropriate way, so that its fixed loci consist in isolated \textit{fixed points}. Once can show that the results are independent of this modification  \cite{Cirafici:2008sn}. Mathematically this procedure is equivalent to using an equivariant (virtual) localization formula to compute the Donaldson-Thomas invariants directly, as we will see momentarily.

As we have just explained, we will work equivariantly with respect to a certain toric action. To this end, it is useful to lift the natural toric action of $\mathbb{C}^3$ to the instanton moduli space. Explicitly on the coordinates of $\mathbb{C} [z_1 , z_2 , z_3]$, the natural torus $\mathbb{T}^3$ acts as $z_\alpha \longrightarrow \mathrm{e}^{\mathrm{i} \epsilon_\alpha} z_\alpha$. We define the following transformation rules under the full group $U(n) \times U(r) \times \mathbb{T}^3$
\begin{eqnarray} \label{transfbosons}
B_\alpha ~&\longmapsto~& \e^{- \ii \epsilon_\alpha}\, g_{U(n)} \, B_\alpha \, g_{U(n)}^{\dagger} \ , \nonumber\\[4pt]
\varphi~&\longmapsto~& \e^{- \ii (\epsilon_1 +
\epsilon_2 + \epsilon_3)} \,  g_{U(n)} \, \varphi \,
g_{U(n)}^{\dagger} \ , \nonumber\\[4pt]
I ~&\longmapsto~& g_{U(n)}\, I\, g_{U(r)}^{\dagger}\ .
\end{eqnarray}
The above field content is constrained by the quiver quantum mechanics bosonic field equations
\begin{eqnarray}
\mathcal{E}_{\alpha} \,&:&\, [B_\alpha , B_\beta] + \sum_{\gamma=1}^3\, \epsilon_{\alpha\beta\gamma} \,\big[B_\gamma^{\dagger}
\,,\, \varphi\big] = 0 \ , \nonumber\\[4pt]
\mathcal{E}_\lambda  \,&:&\, \sum_{\alpha=1}^{3}\, \big[B_\alpha \,,\,
B_\alpha^{\dagger}\,\big] + \big[\varphi \,,\, \varphi^{\dagger}\,\big] + I\,
I^{\dagger} = \varsigma \ , \nonumber\\[4pt] \mathcal{E}_I \,&:&\,
I^{\dagger} \,\varphi = 0 \ .
\label{quiverdefeqs}\end{eqnarray}
Here $\varsigma$ is a Fayet-Iliopoulos parameter.

The topological quiver quantum mechanics is constructed out of these equations. To this end one defines a BRST operator $\mathcal{Q}$ which acts as
\begin{equation}
\begin{array}{rlllrl}
\mathcal{Q} \, B_\alpha &=& \psi_\alpha & \quad \mbox{and} & \quad \mathcal{Q} \, \psi_\alpha& = \ [\phi , B_\alpha] -
\epsilon_\alpha\,
B_\alpha \ , \\[4pt] \mathcal{Q} \, \varphi& =& \xi & \quad \mbox{and} & \quad
\mathcal{Q} \, \xi &= \ [\phi ,\varphi] - (\epsilon_1 + \epsilon_2 + \epsilon_3) \varphi \ , \\[4pt]
\mathcal{Q} \, I &=& \varrho & \quad \mbox{and} & \quad \mathcal{Q} \, \varrho& = \
\phi \,I - I \,\mathbf{a} \ ,
\end{array}
\label{BRSTmatrices}
\end{equation}
where $\mathbf{a} = \mathrm{diag} (a_1 , \dots , a_r)$ parametrizes the Cartan subalgebra $\mathfrak{u}(1)^{\oplus r}$ and $\phi$ is the generator of $U(n)$ gauge transformations. We omit the details of the construction of the full quiver quantum mechanics: one proceeds by introducing Fermi multiplets corresponding to the anti-ghosts and auxiliary fields with the same transformation properties as the equations (\ref{quiverdefeqs}), as well as the gauge multiplet necessary to close the BRST algebra. The construction is such that the partition  function of the quiver quantum mechanics localizes onto the fixed points of the BRST-charge.

These fixed points can be classified explicitly in terms of certain combinatorial arrangements, called plane partitions \cite{Cirafici:2008sn}. A plane partition is a three dimensional Young diagram, which can be obtained by an ordinary Young diagram $\lambda$, by defining a ``box piling function'' $\pi \, : \, \lambda \longrightarrow \mathbb{Z}_+$, with the condition that $\pi_{i,j} \ge \pi_{i+m,j+n}$ with $n,m \in \mathbb{Z}_{\ge 0}$. Equivalently a plane partition can be defined as the complement of a certain ideal. Define the monomial ideal in the polynomial ring $\mathbb{C} [z_1 , z_2 , z_3]$
\begin{equation}
I_m (z_1,z_2,z_3) = \mathbb{C} \langle z_1^{m_1} z_2^{m_2} z_3^{m_3} \ \vert \ m_1+m_2+m_3 \ge m
\rangle
\end{equation}
It can be shown that this ideal has codimension $n= \frac16 m (m+1) (m+2)$. The associated plane partition
\begin{equation}
\pi_m = \{ (m_1 , m_2 , m_3) \in \mathbb{Z}^3_{\ge 0} \ \vert \ z_1^{m_1} z_2^{m_2} z_3^{m_3} \notin I_m \}
\end{equation}
has $|\pi_m| = n$ boxes.

More precisely a fixed point of the BRST charge $\mathcal{Q}$ correspond to a vector $\vec{\pi} = (\pi_1 , \cdots , \pi_r)$ of plane partitions, with $| \vec{\pi} | = \sum_l |\pi_l| = k$ boxes. From the above correspondence between plane partitions and ideals, each plane partition $\pi_i$ describes geometrically a $\mathbb{T}^3$-fixed ideal sheaf $\mathcal{I}_{\pi_i}$ with support on a $\mathbb{T}^3$ invariant zero dimensional subscheme in $\mathbb{C}^3$. A fixed point $\vec{\pi}$ corresponds to the sheaf $\mathcal{E}_{\vec \pi} = \mathcal{I}_{\pi_1} \oplus \cdots \oplus \mathcal{I}_{\pi_r}$.

The contribution of each fixed point is obtained by linearizing the equations (\ref{quiverdefeqs}) and performing the resulting gaussian integrals; the result has the form of a ratio of determinants. This is equivalent to compute the integrals over the moduli spaces directly using virtual localization. Virtual localization is a generalization of the usual localization formulae to the case where the integration domain is not a manifold but has a virtual fundamental class (which is typically the case for moduli spaces which arise in physics). The virtual localization formula has the form of a sum over fixed points, each one weighted by the Euler class of the virtual tangent space
\begin{equation} \label{DTlocaformula}
{\tt DT}_{n , r} \left( \mathbb{C}^3 \right) = \int_{[ \mathcal{M}^{\rm inst}_{n,0;r}(\mathbb{C}^3) ]^{\rm vir}} \ 1 = \sum_{[\mathcal{E}_{\vec{\pi}}] \in \mathcal{M}^{\rm inst}_{n,0;r}(\mathbb{C}^3)^{\mathbb{T}^3 \times U(1)^r} } \ \frac{1}{\mathrm{eul} \left( T_{\vec{\pi}}^{\mathrm{vir}}  \mathcal{M}^{\rm inst}_{n,0;r}(\mathbb{C}^3)  \right) } \ .
\end{equation}
The virtual tangent space is defined as 
\begin{equation}
T_{\vec{\pi}}^{\mathrm{vir}}  \mathcal{M}^{\rm inst}_{n,0;r}(\mathbb{C}^3) = T_{\vec{\pi}}\mathcal{M}^{\rm inst}_{n,0;r}(\mathbb{C}^3) \ominus ( \mathcal{N}_{n,0;r})_{\vec{\pi}} = \mathrm{Ext}^1 \left( \mathcal{E}_{\vec{\pi}} , \mathcal{E}_{\vec{\pi}} \right)  \ominus \mathrm{Ext}^2 \left( \mathcal{E}_{\vec{\pi}} , \mathcal{E}_{\vec{\pi}} \right) \ .
\end{equation}
Note that (\ref{DTlocaformula}) has precisely the form of the naive integrals in (\ref{Zdt}): the Euler class of the virtual tangent space is by definition the ration between the Euler classes of the tangent and obstruction bundles. This is what one would get evaluating  (\ref{Zdt}) using naive localization over smooth manifolds with the specific integration measure dictated by topological six dimensional Yang-Mills! As we have promised, the partition function of the quiver quantum mechanics compute directly the BPS invariants.

The Euler class of the virtual tangent space can be computed from a quiver quantum mechanics version of the instanton deformation complex (\ref{defcomplex}). To this end, we decompose the vector spaces $V$ and $W$ in the representation ring of $\mathbb{T}^3 \times U(1)^r$ as
\begin{eqnarray}
V_{\vec\pi} = \sum_{l=1}^r \,e_l~ \sum_{(n_1,n_2,n_3)\in \pi_l}\,
t_1^{n_1-1} \,t_2^{n_2-1}\,t_3^{n_3-1} \qquad \mbox{and} \qquad W_{\vec\pi} =
\sum_{l=1}^r \,e_l \ .
\label{decompos}
\end{eqnarray}
Here we have introduced $e_l= \e^{\ii a_l}$ and $t_\alpha = \e^{\ii \epsilon_{\alpha}}$ for $\alpha=1,2,3$. To keep track of the toric action it is useful to introduce the $\mathbb{T}^3$ module $Q \simeq \mathbb{C}^3$ generated by $t_\alpha^{-1} =  \e^{-\ii \epsilon_{\alpha}}$. From (\ref{transfbosons}) we see that for given a fixed point $\vec{\pi}$,  $(B_1 , B_2 , B_3) \in \mathrm{End}_{\mathbb{C}} (V_{\vec{\pi}}) \otimes Q$ and $I \in \mathrm{Hom}_{\mathbb{C}} (W_{\vec \pi} , V_{\vec \pi})$. To study the local geometry of the moduli space around this fixed point, we define the instanton deformation complex
\begin{equation} \label{adhmdefcomplexC}
\xymatrix{
  \mathrm{Hom}_\mathbb{C} (V_{\vec\pi} , V_{\vec\pi})
   \quad\ar[r]^{\!\!\!\!\!\!\!\!\!\!\!\!\!\!\!\!\sigma} &\quad
   {\begin{matrix} \mathrm{Hom}_\mathbb{C} (V_{\vec\pi} , V_{\vec\pi} \otimes Q )
   \\ \oplus \\
   \mathrm{Hom}_\mathbb{C} (W_{\vec\pi} , V_{\vec\pi}) \\ \oplus  \\ \mathrm{Hom}_\mathbb{C} (V_{\vec\pi} ,
   V_{\vec\pi}  \otimes \bigwedge^3 
   Q) \end{matrix}}\quad \ar[r]^{\tau} & \quad
   {\begin{matrix} \mathrm{Hom}_\mathbb{C} (V_{\vec\pi} , V_{\vec\pi}  \otimes \bigwedge^2
       Q) \\ \oplus \\ 
       \mathrm{Hom}_\mathbb{C} (V_{\vec\pi},W_{\vec\pi} \otimes \bigwedge^3 Q)
   \end{matrix}}
} \ .
\end{equation}
Here $\tau$ is the linearization of the equations $\mathcal{E}_\alpha$ and $\mathcal{E}_I$, while $\sigma$ is an infinitesimal complex gauge transformation. The first cohomology of this complex is a model for the tangent space to the moduli space at the fixed point $\vec{\pi}$ and its second cohomology is a model for the normal bundle. We assume that the cohomology at order zero, which corresponds to reducible connections, vanishes. Therefore the equivariant index of the complex (\ref{adhmdefcomplexC}) computes the virtual sum $\mathrm{Ext}^1 \ominus \mathrm{Ext}^2$. The equivariant index can be written down explicitly at a fixed point $\vec \pi$ in terms of the characters of the representations as
\begin{equation} \label{character}
\mathrm{ch}_{\mathbb{T}^3 \times U(1)^r} \left(  T_{\vec{\pi}}^{\mathrm{vir}}  \mathcal{M}^{\rm inst}_{n,0;r}(\mathbb{C}^3)  \right) = 
W_{\vec\pi}^\vee \otimes V_{\vec\pi} -
{V}_{\vec\pi}^\vee \otimes W_{\vec\pi} + (1-t_1)\, (1-t_2)\,
(1-t_3)~ {V}^\vee_{\vec\pi} \otimes V_{\vec\pi} \ ,
\end{equation}
where we have used the fact that $\mathbb{C}^3$ is (trivially) Calabi-Yau to set $\epsilon_1 + \epsilon_2 + \epsilon_3 = 0$. From the equivariant character it is straightforward to obtain the equivariant Euler class of (\ref{DTlocaformula}) as the equivariant top Chern class. It turns out that the result is just a sign, and in particular is independent on the equivariant parameters $\epsilon_i$ and $a_l$ (although this dependence would be reintroduced were we to drop the condition $\epsilon_1 + \epsilon_2 + \epsilon_3 = 0$). In this way we obtain an explicit presentation of the formal partition function (\ref{Zdt}):
\begin{equation}
Z^{\mathbb{C}^3}_{\mathrm{gauge}} (q;r) = \sum_{|\vec{\pi}|} \, q^{|\vec{\pi}|} \ 
\frac{\mathrm{eul}(\mathcal{N}_{n,0;r})_{\vec{\pi}}}{\mathrm{eul}\big(T_{\vec{\pi}}\mathcal{M}^{\rm inst}_{n,0;r}(\mathbb{C}^3)\big)} =   \sum_{\vec{\pi}} \ (-1)^{r |\vec{\pi}|} \ q^{|\vec{\pi}|} \ .
\end{equation}
Note that this partition function is explicitly defined only in the Coulomb branch; to compute truly nonabelian invariants one would have to impose an appropriate stability condition on the moduli space of coherent sheaves and then only retain the relevant fixed points. 

\section{Localization on toric varieties} \label{localization}

The construction we have just presented was explicitly based on the properties of $\mathbb{C}^3$ and one could wonder how general it is. It turns out that the same construction can be extended to a broad class of varieties called toric varieties. The construction is straightforward but somewhat aside from the main themes of this survey and therefore we will only sketch the main points. 

The geometry of a toric Calabi-Yau threefold $X$ can be combinatorially encoded in a trivalent graph $\Delta (X)$, known as a toric graph. Roughly speaking a toric threefold is a variety which contains an algebraic torus $\mathbb{T}^3$ as an open dense subset. This torus naturally acts on the whole variety. The toric action is Hamiltonian and the graph $\Delta (X)$ is the image of $X$ under its moment map. The geometry of $X$ is encoded in $\Delta (X)$ as follows: the trivalent vertices $f$ of $\Delta (X)$ are in correspondence with the fixed points of the toric action on $X$ and each fixed point is at the origin of a toric invariant open $\mathbb{C}^3$ chart. The edges of the graph $\Delta (X)$ correspond to $\mathbb{T}^3$-invariant projective lines $\mathbb{P}^1$, such that two fixed points $f_N$ and $f_S$ can be identified respectively with the north and south pole of a $\mathbb{P}^1$.

A gauge theory on a toric threefold localizes onto contribution coming from toric invariant configurations \cite{Iqbal:2003ds,Cirafici:2008sn}. These are point-like instantons located at the vertices of $f$ and extended instantons spread over the toric lines $\mathbb{P}^1$. Combinatorically these are represented by three-dimensional Young diagrams $\pi$ associated with the vertices of $\Delta (X)$ and ordinary two-dimensional Young diagrams $\lambda$ associated with the each edge $e$ of $\Delta (X)$ and represent four-dimensional instantons fibered over the $\mathbb{P}^1$.

The result is that the partition function of a rank $r$ gauge theory in the
Coulomb branch is given by~\cite{Cirafici:2008sn}
\begin{align}
{Z}_{\rm gauge}^{X}(q,Q;r) = & \sum_{\vec\pi_v,\vec\lambda_e} \,
(-1)^{r\, D\{\vec\pi_v,\vec\lambda_e\}} \,
 q^{D\{\vec\pi_v,\vec\lambda_e\}} \cr & \times \prod_{ {\rm edges}\ e} \,
(-1)^{\sum_{l,l'=1}^r\, |\lambda_{e,l}|\, |\lambda_{e,l'}|\, m_{e,1}} \,
Q_e^{\sum_{l=1}^r \, |\lambda_{l,e}|} \ ,
\end{align}
where 
\begin{equation}
D\{\vec\pi_v,\vec\lambda_e\}= \sum_{{\rm vertices}\ v} \ \sum_{l=1}^r \, |\pi_{v,l}|
+ \sum_{{\rm edges}\ e} \  \sum_{l=1}^r \ \sum_{(i,j) \in
  \lambda_{e,l}} \, \big(
m_{e,1}\,  (i-1) + m_{e,2} \, (j-1) + 1 \big)
\end{equation}
Here the pair of integers $(m_{e,1},m_{e,2})$ specify the normal bundles
over the projective lines associated with the edges $e$ of the graph
$\Delta$.

Similar formulas exists for four-dimensional gauge theories on toric
surfaces~\cite{neklocal,Gasparim:2008ri,Cirafici:2009ga,Cirafici:2012qc}.

\section{Noncommutative Crepant Resolutions} \label{noncommutative}

Now we will address Donaldson-Thomas theory in another chamber, sometimes called the noncommutative crepant resolution chamber. This chamber is in a sense ``non-geometric", the target space description cannot be understood in terms of geometrical terms but requires a more algebraic perspective. Geometrically this chamber is associated with a singular threefold, where the manifolds structure breaks down at the singularity. In many cases a smooth local Calabi-Yau threefold can be described as the moduli space of representations of a certain quiver $\mathsf{Q}$, where all the vector spaces are one dimensional. Indeed this is precisely the crepant resolution of an abelian orbifold singularity $\mathbb{C}^3 / \Gamma$, with $\Gamma$ a subgroup of $SL(3,\mathbb{C})$, given by the $\Gamma$-Hilbert scheme $\mathrm{Hilb}^{\Gamma} (\mathbb{C}^3)$ which parametrizes $\Gamma$-invariants schemes \cite{ItoNaka}.

However under certain circumstances, the path algebra $\mathsf{A}$ itself of the quiver, can be understood as a resolution of the singularity \cite{vandenbergh}. This is a particular instance of a broader program, known as Noncommutative Algebraic Geometry, where the usual local models of Algebraic Geometry, consisting in commutative rings or algebras, are replaced by noncommutative structures, algebraic or categorical. We will not discuss the general theory, but proceed by examples. 

Consider for example the conifold singularity. It can be described as the locus $z_1 z_2 - z_3 z_4 = 0$ in $\mathbb{C}^4$. Its crepant resolution is called the resolved conifold and is the total space of the holomorphic bundle $\mathcal{O}_{\mathbb{P}^1} (-1) \oplus \mathcal{O}_{\mathbb{P}^1} (-1) \longrightarrow \mathbb{P}^1$. Equivalently the deformation can be described as the locus $z_1 z_2 - z_3 z_4 = t$, with $t$ representing the area of the $\mathbb{P}^1$ replacing the singularity at the origin. 

To discuss the noncommutative crepant resolution, the relevant quiver is the Klebanov-Witten quiver~\cite{Klebanov:1998hh}
\begin{equation}
\begin{xy}
\xymatrix@C=30mm{
  \circ \ \ar@/_1pc/[r]|{ \ a_1 \ } \ar@/_2pc/[r]|{ \ a_2 \ } &  \
  \ar@/_1pc/[l]|{ \ b_1 \ } \ar@/_2pc/[l]|{ \ b_2 \ } \ \bullet
}
\end{xy}
\end{equation}
with superpotential
\begin{equation}
{\sf W} = a_1 \, b_1 \,a_2 \,b_2 - a_1\, b_2 \,a_2\, b_1 \ .
\end{equation} 
We can describe the path algebra explicitly as
\begin{equation}
{\sf A} = \mathbb{C} [e_\circ  ,  e_\bullet] \langle a_1 , a_2 , b_1 , b_2 \rangle
\, \big/ \, \langle b_1 \,a_i\, b_2 - b_2\, a_i\, b_1 \,,\, a_1\,
b_i\, a_2 - a_2\, b_i \,a_1 \ | \ i = 1,2 \rangle \ .
\end{equation}
where $e_\circ$ and $e_\bullet$ are the trivial paths of length zero at the nodes $\circ$ and $\bullet$. 

The centre $\mathsf{Z} (\mathsf{A})$ of this algebra is generated by the elements
\begin{eqnarray}
z_1 &=& a_1\, b_1 + b_1\, a_1 \ , \nonumber \\[4pt]
z_2 &=& a_2 \,b_2 + b_2\, a_2 \ , \nonumber \\[4pt]
z_3 &=& a_1 \,b_2 + b_2\, a_1 \ , \nonumber \\[4pt]
z_4 &=& a_2 \,b_1 + b_1\, a_2 \ ,
\end{eqnarray}
and hence
\begin{equation}
{ \mathsf{Z}(\mathsf{A})} = \mathbb{C} [z_1 , z_2 , z_3 , z_4] \, \big/ \, \left( z_1 \,z_2 - z_3 \,z_4 \right)
\end{equation}
In other words the path algebra of the conifold quivers contains the nodal singularity of the conifold as its center. This is our first example of noncommutative resolution.

Noncommutative crepant resolutions admit BPS invariants which corresponds to bound states of D-branes, the noncommutative Donaldson-Thomas invariants \cite{szendroi,reineke,Ooguri:2008yb,Cirafici:2010bd,Cirafici:2011cd}. Roughly speaking one can construct a supersymmetric quiver quantum mechanics with superpotential out of the data of the conifold quiver; this quantum mechanics describes the effective field theory of a system of $D2-D0$ branes on the conifold. To properly have Donaldson-Thomas type invariants, we also need a magnetic charge; in this case we can add a single noncompact $D6$ brane wrapping the whole of the conifold geometry. The effect of this modification is that now the conifold quiver is framed as
\begin{equation}
\begin{xy}
\xymatrix@C=30mm{
\star \ \ar@{.>}@//[dr]|< < < < < < < <{ \ a_{\star} \ } \\   & \ \circ  \ \ar@/_1pc/[r]|{
  \ a_1 \ } \ar@/_2pc/[r]|{ \ a_2 \ } &  \ \ar@/_1pc/[l]|{ \ b_1 \ }
\ar@/_2pc/[l]|{ \ b_2 \ } \ \bullet 
}
\end{xy}
\end{equation}
Noncommutative Donaldson-Thomas invariants are defined as enumerative invariants associated with the moduli space of framed representations of this quiver. To define this moduli space, we start from the representation space
\begin{equation}
Rep (\widehat{\mathsf{Q}} , \circ  ) = \bigoplus_{(v\longrightarrow w) \in
  {\mathsf{Q}}_1}\,  \mathrm{Hom}_\mathbb{C} (V_{v} , V_{w}) \ \oplus \ \mathrm{Hom}_\mathbb{C} (V_\circ ,
\mathbb{C}) \ ,
\end{equation}
which explicitly depends on the choice of node $\circ$ of the quiver, which is framed. Let $Rep (\widehat{\mathsf{Q}} , \circ ; \mathsf{W})$ be the subscheme of $Rep (\widehat{\mathsf{Q}} , \circ )$ cut out by the superpotential equations $\partial_a \mathsf{W} = 0$. The relevant moduli space is the smooth Artin stack\footnote{In this note we will refrain to discuss stacks; the only examples of algebraic or Artin stacks we will encounter are those obtained from the quotient of a scheme $S$ (which is an algebraic stack on its own right) by an algebraic group. In general algebraic stacks do not have a well defined notion of integration; our case is an exception since it is defined as the vanishing locus $\partial_a \mathsf{W} = 0$ where $\mathsf{W}$ is gauge invariant. These conditions essentially define a symmetric perfect obstruction theory and therefore a virtual fundamental cycle.}
\begin{equation}
{\mathcal{M}}_{n_0,n_1}(\widehat{\mathsf{Q}}) = \big[ Rep (\widehat{\mathsf{Q}} , \circ ; \mathsf{W} ) \,
\big/ \, GL(n_0,\mathbb{C})\times GL(n_1,\mathbb{C}) \big] \ ,
\end{equation}
Noncommutative Donaldson-Thomas invariants can now be defined as (weighted) Euler characteristics of the moduli spaces ${\mathcal{M}}_{n_0,n_1}(\hat{\mathsf{Q}})$ and studied explicitly. Indeed the invariants were computed using equivariant localization and the problem admits a purely combinatorial solution \cite{szendroi}. In the following we will discuss these issues from a slightly different perspective for more general singularities. We refer the reader to \cite{Jafferis:2008uf} for a more in-depth discussion of BPS states on the conifold.

\section{Instantons on $[ \mathbb{C}^3 / \Gamma ]$ and McKay quivers} \label{instantons}

We will now consider singularities of the form $\mathbb{C}^3 / \Gamma$ where $\Gamma$ is a finite subgroup of $SL (3 , \mathbb{C})$. In this case the relevant quiver is the so-called McKay quiver $\mathsf{Q}_\Gamma$, which is constructed out of the representation data of the finite group $\Gamma$. This quiver has a node for each irreducible one-dimensional representation $\rho_a$ of $\Gamma$. We will denote by $\hat{\Gamma}$ the group of  such representations. The arrow structure and the relations are determined by the tensor product decomposition
\begin{equation} \label{tensordecomp}
\mbox{$\bigwedge^i$}\, Q \otimes \rho_a =
\bigoplus_{b\in\widehat\Gamma}\, a^{(i)}_{ba}\, \rho_b \qquad
\mbox{with} \quad a^{(i)}_{ba}=\dim_\mathbb{C} \mathrm{Hom}_\Gamma\big(\rho_b
\,,\,\mbox{$\bigwedge^i$}\, Q \otimes \rho_a \big)
\end{equation}
Here $Q = \rho_{a_1} \oplus \rho_{a_2} \oplus \rho_{a_3}$ is the fundamental three-dimensional representation of $\Gamma$, corresponding to the action of $\Gamma$ on $\mathbb{C}^3$ with weights $a_\alpha$, $\alpha=1,2,3$ (such that $a_1+a_2+a_3 = 0$ since $\Gamma$ is a subgroup of $SL(3,\mathbb{C}^3)$). In particular one can show that for a Calabi-Yau singularity $a^{(1)}_{ba} = a^{(2)}_{ab}$ and $a^{(3)}_{ab}= \delta_{ab}$. The McKay quiver has $a^{(1)}_{ab}$ arrows going from node $\rho_b$ to node $\rho_a$. 

As an example, consider for example the orbifold $\mathbb{C}^3 / \mathbb{Z}_3$. We let the generator $g$ of $\mathbb{Z}_3$ act on $\mathbb{C}^3$ as $g (z_1 , z_2 , z_3) = ( \e^{2 \pi \ii/3} z_1, \e^{2 \pi \ii/3} z_2, \e^{2 \pi \ii/3}z_3)$. The relevant quiver is
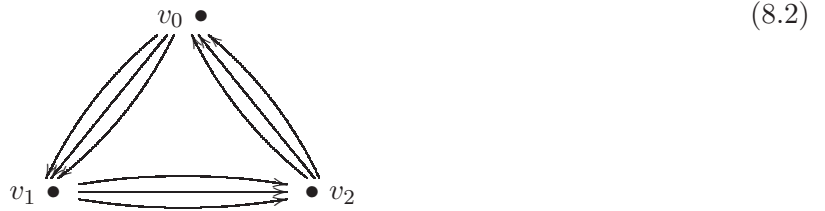
\begin{equation}
\vspace{4pt}
\begin{xy}
\xymatrix@C=8mm{
& \ v_0 \ \bullet \ \ar@/^/[ddl] \ar@/_0.5pc/[ddl] \ar@//[ddl]  & \\
& & \\
v_1 \ \bullet \ \ar@//[rr] \ar@/^/[rr]  \ar@/_/[rr]   & &  \ \bullet \ v_2  \ar@/^/[uul] \ar@/_0.5pc/[uul] \ar@//[uul] 
}
\end{xy}
\vspace{4pt}
\label{quiverC3Z3}\end{equation}
with weights $a_\alpha=1$ for $\alpha=1,2,3$, i.e. in this case $Q = \rho_1 \oplus \rho_1 \oplus \rho_1$. In particular we see from (\ref{tensordecomp}) that
\begin{equation}
 a_{ab}^{(1)}= \left( \begin{matrix} 0 & 0 & 3 \\ 3 & 0 & 0 \\  0 & 3 & 0 \end{matrix} \right)
 \qquad \mbox{and} \qquad 
 a_{ab}^{(2)} = \left( \begin{matrix} 0 & 3 & 0 \\ 0 & 0 & 3 \\  3 & 0 & 0 \end{matrix} \right)
\label{C3Z3matrices}\end{equation}

A representation of the McKay quiver $\mathsf{Q}_\Gamma$ has a natural $\Gamma$-module structure. From the individual vector spaces based at the nodes we construct $V = \bigoplus_{a\in \hat{\Gamma}}\, V_a \otimes \rho^\vee_a$ (here $\rho^\vee$ is the conjugate representation). The linear maps between the nodes can be encoded in $B \in \mathrm{Hom}_{\Gamma} (V , Q \otimes V)$, or equivalently they decompose as $B = \bigoplus_{a\in \hat{\Gamma}}\, \big( B_1^{(a)} \,,\, B_2^{(a)} \,,\, B_3^{(a)} \big)$ where $B_\alpha^{(a)} \in \mathrm{Hom}_\mathbb{C} (V_{a}, V_{a+ a_\alpha})$. With this notation, the ideal of relations $\langle \mathsf{R}_\Gamma \rangle$  of the McKay quiver corresponds to the following matrix equations
\begin{equation}
B_\beta^{(a + a_\alpha)} \ B_\alpha^{(a)} = B_\alpha^{(a + a_\beta)} \ B_\beta^{(a)} \qquad \mbox{for} \quad
\alpha,\beta=1,2,3 \ .
\label{ADHMorb}\end{equation}
The path algebra $\mathsf{A}_\Gamma = \mathbb{C} \mathsf{Q}_\Gamma / \langle \mathsf{R}_\Gamma \rangle$ is a noncommutative crepant resolution of the singularity $\mathbb{C} / \Gamma$. 

Noncommutative Donaldson-Thomas invariants in the noncommutative crepant resolution chamber can still be understood from a gauge theory perspective. The gauge theory at hand is similar to the one discussed above, but it has to be suitably deformed. In particular the model is a topologically twisted version of a noncommutative field theory defined on the algebraic stack $[\mathbb{C}^3 / \Gamma]$. This class of theories were dubbed ``stacky'' gauge theories, and can be simply thought of as field theories on $\mathbb{C}^3$ whose observables  are $\Gamma$-equivariant quantities \cite{Cirafici:2010bd,Cirafici:2011cd}. In particular these theories admit generalized instanton solutions whose moduli spaces can be constructed rather explicitly via the McKay correspondence, generalizing the celebrated Kronheimer-Nakajima constructions of instantons on ALE spaces \cite{KN} (and the D-brane description of \cite{Douglas:1996sw}). 

The relevant instanton moduli spaces can be described in terms of the representations of the framed McKay quiver associated with the singularity $\mathbb{C}^3 / \Gamma$. The vector spaces associated to each node of $\mathsf{Q}_\Gamma$ are those which enter in the isotypical decomposition $V = \bigoplus_{a\in \hat{\Gamma}}\, V_a \otimes \rho^\vee_a$, where $\mathrm{dim} V = k$ represents the instanton number, while the individual dimensions $\mathrm{dim} V_a = k_a$ correspond to instanton configurations which transform in the irreducible representation $\rho_a$. Similarly the information about the framing nodes can be encoded in the decomposition 
$W=\bigoplus_{a\in\widehat\Gamma}\, W_a\otimes \rho_a^\vee$: physically the framing nodes label boundary conditions at infinity. At infinity the gauge field is a flat connection labelled by a representation $\rho$ of $\Gamma$, and the dimensions $\mathrm{dim}_{\mathbb{C}} \, W_a = r_a$ label the multiplicities of the decomposition of $\rho$ into the irreducible representations $\rho_a$, provided that $\sum_{a\in\hat{\Gamma}}\, r_a =r $. Finally the framing nodes are connected with the quiver $\mathsf{Q}_\Gamma$ by linear maps $I\in\mathrm{Hom}_\Gamma(W,V)$, which decompose as $I = \bigoplus_{a\in\hat{\Gamma}}\, I^{(a)}$ with $I^{(a)}\in \mathrm{Hom}_\mathbb{C} (W_a,V_a)$.

Overall this construction give a correspondence between a gauge field sheaf $\mathcal{E}$ with prescribed boundary conditions at infinity and a representation of the framed McKay quiver. This relation can be regarded more geometrically using the McKay correspondence. Simply put the McKay correspondence connects smooth geometry of the canonical crepant resolution of $\mathbb{C}^3 / \Gamma$ given by $X=\mathrm{Hilb}^\Gamma (\mathbb{C}^3)$ with representation theory data associated with $\Gamma$. On the resolution $X$ there is a canonical integral basis for the Grothendieck group $K(X)$ of vector bundles, given by the tautological bundles. To define these, consider the universal scheme $\mathcal{Z} \subset X \times \mathbb{C}^3$, with correspondence diagram
\begin{equation}
\xymatrix@=10mm{
  & \mathcal{Z} \ar[ld]_{p_1}\ar[rd]^{p_2}& \\
  X & & \mathbb{C}^3
}
\end{equation}
Then we set $\mathcal{R} := p_{1*} \mathcal{O}_{\mathcal{Z}}$ and define the tautological bundles via the decomposition $\mathcal{R} = \bigoplus_{a\in\widehat\Gamma}\, \mathcal{R}_a \otimes \rho_a
$. To this basis we associate the dual basis $\mathcal{S}_a$ of $K^c (X)$ of compactly supported coherent sheaves. Via the McKay correspondence these basis correspond to the basis $\{ \rho_a \otimes \mathcal{O}_{\mathbb{C}^3} \}_{a \in \hat{\Gamma}}$ and $\{ \rho_a \otimes \mathcal{O}_0 \}_{a \in \hat{\Gamma}}$ of $K_\Gamma (\mathbb{C}^3)$ and $K_{\Gamma}^c (\mathbb{C}^3)$ the Grothendieck groups of $\Gamma$-equivariant sheaves and $\Gamma$-equivariant sheaves with compact support on $\mathbb{C}^3$ respectively (here $\mathcal{O}_0$ is the skyscraper sheaf at the origin). We refer the reader to \cite{Cirafici:2012qc} for a much more detailed account of the uses of the McKay correspondence and its generalization within the present context. Using these data, the Chern character of the gauge field sheaf $\mathcal{E}$ can  be written as
\begin{eqnarray} \label{chE}
\mathrm{ch} (\mathcal{E}) &=& - \mathrm{ch} \Big( \big( V \otimes \mathcal{R} (-2) \big)^{\Gamma} \Big) +
 \mathrm{ch} \Big(
 \big(\mbox{$V \otimes \bigwedge^2 Q^{\vee}$} \otimes \mathcal{R} (-1) \big)^{\Gamma} \Big)
 \cr & & -\, \mathrm{ch} \Big(
\big(( {V \otimes Q^{\vee} \oplus W}) \otimes \mathcal{R} \big)^{\Gamma}
 \Big) + \mathrm{ch} \Big( \big(  {V} \otimes \mathcal{R} (1) \big)^{\Gamma} \Big) \ .
\end{eqnarray}

The McKay correspondence allows us to extend the formalism we have described to study BPS invariants on $\mathbb{C}^3$ to the case at hand. Indeed thanks to the fact that $\Gamma$ is a subgroup of the torus group $\mathbb{T}^3$ the formalism extends almost verbatim. The relevant quiver quantum mechanics has superpotential
\begin{equation}
\mathsf{W}_{\Gamma}= \sum_{a\in\hat{\Gamma}}\, 
B_1^{(a+a_2+a_3)}\,\Big( B_2^{(a+a_3)}\, B_3^{(a)}-B_3^{(a+a_2)}\,
  B_2^{(a)} \Big) \ .
\end{equation}
As before the quiver quantum mechanics localizes onto fixed points of its BRST operator. In particular, since the $\Gamma$ action and the $\mathbb{T}^3$ action commute, the fixed point set is the same as in the $\mathbb{C}^3$ case, $r$-vectors of plane partitions $\vec{\pi} = (\pi_1 , \cdots , \pi_r)$, the only difference being that one has to carefully take into account the $\Gamma$ action. 

The moduli space associated with the quiver quantum mechanics is the quotient stack
\begin{equation}
\mathcal{M}^{\Gamma}_{\mathbf{k},\mathbf{N}} = [ Rep (\mathsf{\hat{Q}}_{\Gamma} ; \mathsf{W}_{\Gamma}) / \prod_{r\in\widehat\Gamma}\, GL(k_r,\mathbb{C}) ]
\end{equation}
where $Rep (\mathsf{\hat{Q}}_{\Gamma} ; \mathsf{W})$ is the subset of
\begin{equation}
 \mathrm{Hom}_{\Gamma} (V , Q \otimes V) \ 
 \oplus \ \mathrm{Hom}_{\Gamma} (V , \mbox{$\bigwedge^3$} Q \otimes V) \ 
\oplus \ \mathrm{Hom}_{\Gamma} (W , V) \ 
\end{equation}
cut out by the equations derived from the superpotential $\mathsf{W}_{\Gamma}$. A local model of the moduli space around a fixed point is given by the deformation complex
\begin{equation} \label{equivdefcomplex}
\xymatrix{
  \mathrm{Hom}_{\Gamma}  (V_{\vec\pi} , V_{\vec\pi})
   \quad\ar[r] &\quad
   {\begin{matrix}  \mathrm{Hom}_{\Gamma} (V_{\vec\pi} , V_{\vec\pi} \otimes Q )
   \\ \oplus \\
    \mathrm{Hom}_{\Gamma} (W_{\vec\pi} , V_{\vec\pi}) \\ \oplus  \\  \mathrm{Hom}_{\Gamma} (V_{\vec\pi} ,
   V_{\vec\pi}  \otimes \bigwedge^3 
   Q) \end{matrix}}\quad \ar[r] & \quad
   {\begin{matrix}  \mathrm{Hom}_{\Gamma} (V_{\vec\pi} , V_{\vec\pi}  \otimes \bigwedge^2
       Q) \\ \oplus \\ 
        \mathrm{Hom}_{\Gamma} (V_{\vec\pi},W_{\vec\pi} \otimes \bigwedge^3 Q)
   \end{matrix}}
}
\end{equation}
from which we can extract the character at the fixed points
\begin{equation} \label{orbcharacter}
\mathrm{Ch}_{\vec\pi}^\Gamma(t_1,t_2,t_3)= \big( W_{\vec\pi}^\vee \otimes V_{\vec\pi} -
{V}_{\vec\pi}^\vee \otimes W_{\vec\pi}+ (1-t_1)\, (1-t_2)\,
(1-t_3) ~ {V}^\vee_{\vec\pi} \otimes V_{\vec\pi} \big)^{\Gamma} \ .
\end{equation}
The vector spaces $V$ and $W$ once again can be decomposed at a fixed point $\vec \pi$ as in (\ref{decompos}). However now each partition carries an action of the group $\Gamma$: the fundamental orbifold representation $Q = \rho_{a_1} \oplus \rho_{a_2} \oplus \rho_{a_3}$ induces an action with weight $a_i$ on each module generator $t_i$, $i=1,2,3$. Therefore each box of the plane partition is associated with a character of $\Gamma$. This action is however ``offset" by the transformation of $e_l = \e^{\ii a_l}$ under $\Gamma$. This transformation encodes the boundary conditions and specifying which eigenvalue $a_l$ transforms in a particular irreducible representations of $\Gamma$ uniquely identifies a superselection sector. This information can be compactly encoded by defining the boundary function $b : \{ 1 , \dots , r\} \longrightarrow \hat{\Gamma}$ which to each sector $l$ with module generator $e_l = \e^{\ii a_l}$ associates the weight $b(l)$ of the corresponding representation of $\Gamma$. Using this notation we can write
\begin{equation}
V_{\vec\pi} =\bigoplus_{l=1}^r ~ \bigoplus_{a\in\widehat{\Gamma}} \, \big( E_l \otimes \rho_{b(l)}^{\vee} \big) \otimes \left( P_{l,a} \otimes \rho_a^{\vee} \right) =\bigoplus_{l=1}^r~ \bigoplus_{a\in\widehat{\Gamma}} \, \big( E_l \otimes P_{l,a} \big) \otimes \rho_{a+b(l)}^{\vee} \, .
\label{VpiGammadecomp}\end{equation}
The modules $P_{l,a}$ correspond to the decomposition of $\sum_{(n_1,n_2,n_3)\in \pi_l}\, t_1^{n_1-1} \,t_2^{n_2-1}\,t_3^{n_3-1} $ as a  $\Gamma$-module. In practice this means that $|\pi_{l,a}| = \mathrm{dim} P_{l,a}
$ is the number of boxes in the $l$-th plane partition in the vector $\vec{\pi}=(\pi_1 , \cdots , \pi_r)$ which transforms in the representation $\rho_r^{\vee}$. In particular the instanton numbers are related with the number of boxes in a partition which transform in a given irreducible representation of $\Gamma$ as
\begin{equation} \label{NAinst}
k_r = \sum_{l=1}^r \, |\pi_{l,a-b(l)}| \ .
\end{equation}
From the character (\ref{orbcharacter}) it is straightforward, if somewhat involved, to derive the contribution of an instanton to the gauge theory fluctuation determinant: it is simply given by a sign $(-1)^{{\mathcal K} (\vec\pi;\mathbf{r})} $, with
\begin{eqnarray}
\mathcal{K}(\vec\pi;\mathbf{r}) &=& \sum_{l=1}^r ~ \sum_{a\in\widehat{\Gamma}}\, |\pi_{l,a}| \ r_{a+b(l)}  
\nonumber \\ &&
- \sum_{l,l'=1}^r \ \sum_{a\in\widehat{\Gamma}}\, |\pi_{l,a}|\, \Big( |\pi_{l',a+b(l)-b(l'\,)-a_1-a_2}| - |\pi_{l',a+b(l)-b(l'\,)-a_1}| \nonumber \\ &&  -\, |\pi_{l',a+b(l)-b(l'\,)-a_2}| + |\pi_{l',a+b(l)-b(l'\,)}| \Big) \ .
\label{instmeasure}\end{eqnarray} 
Finally by collecting all the results, the gauge theory partition function on the stack $[\mathbb{C}^3 / \Gamma]$ has the form
\begin{equation}
Z_{\rm gauge}^{[\mathbb{C}^3 /  \Gamma]}(q,Q;\mathbf{r}) = \sum_{\vec \pi} \,
(-1)^{\mathcal{K}_G(\vec\pi;\mathbf{r})} \, q^{\mathrm{ch}_3(\mathcal{E}_{\vec\pi})} \,
Q^{\mathrm{ch}_2(\mathcal{E}_{\vec\pi})} \ ,
\end{equation}
where the Chern characters can be computed explicitly from (\ref{chE}) using the McKay correspondence, in terms of the characters of the tautological bundles. 

\section{Divisor defects} \label{divisor}

We have argued that using intuition from physics and framed quivers we can get informations about certain enumerative invariants of certain geometries. One could wonder how far can we push this picture; for example if one can conjecturally construct new enumerative invariants. We would like to argue that this seems in principle possible and propose a version of our arguments related to moduli spaces of sheaves with a parabolic structure along a divisor. 

The main idea comes from considering our six dimensional cohomological gauge theory in the presence of defects. Defects have been widely studied in four dimensional field theories, where they are natural order parameters and play a prominent role in the classification of phases of gauge theories. In many cases they can be understood as imposing certain conditions on the fields in the path integral. The six dimensional cohomological gauge theory we have been considering so far is less rich dynamically, but allows the possibility of studying BPS invariants on arbitrary Calabi-Yau threefolds. 

The defects we have in mind are higher dimensional generalization of the surface operators of \cite{Gukov:2006jk,Gukov:2008sn,Gaiotto:2011tf}, and were called \textit{divisor defects} in \cite{Cirafici:2013nja}. As before our gauge theory is defined via a $G$-bundle $\mathcal{E}$ (although in this survey we only consider $G=U(r)$). Let $D$ be a divisor on a Calabi-Yau threefold $X$. Note that $D$ has real co-dimension two in $X$. Defining a divisor defect consists in prescribing a certain singular behavior for the gauge field around $D$. More precisely, locally our space has the form $D \times C$, with $C$ the local fiber of the normal bundle of $D$ in $X$. We pick coordinates on $C$ as $z = r \e^{\ii \theta}$. We require that the gauge field near the defect, that is restricted to $C$, has the form 
\begin{equation} 
A = \alpha \, \mathrm{d} \theta + \cdots \, , 
\end{equation}
where the dots refer to less singular terms. The gauge field has therefore a singularity at the origin of $C$. We extend these arguments globally by requiring that the gauge field has this form at each point of the normal plane to the divisor $D$. The parameter $\alpha$ specifies the type of divisor defect and takes values in the maximal torus $\mathbb{T}_G$ of $G$. Indeed while $\alpha$ naturally takes values in the Cartan sub algebra $\mathfrak{t} = \mathrm{Lie} \mathbb{T}_G$, the correct gauge invariant quantity is the monodromy $\e^{- 2 \pi \alpha}$ valued in $\mathbb{T}_G$. If we introduce the two form $\delta_D$ which is Poincar\'e dual to $D$, the field strength has the form
\begin{equation}
F_A = 2 \pi \alpha \, \delta_D + \cdots \ .
\end{equation}
Because of the singularity, the field theory is naturally defined only on $X \setminus D$. However the bundle $\mathcal{E}$ can be extended on the whole of $X$ albeit in a non unique fashion; indeed extensions of $\mathcal{E}$ over $X$ are in correspondence with lifts of $\alpha$ from $\mathbb{T}_G$ to $\mathfrak{t} = \mathrm{Lie} \, \mathbb{T}_G$. Extensions are mapped into each other by gauge transformations $(r , \theta) \longrightarrow \e^{\theta \, u}$, with $u \in \mathfrak{t}$ with the property $\e^{2 \pi u} = 1$. Note that these gauge transformations are trivial in $\mathbb{T}_G$. In other words, while $\mathcal{E}$ cannot be extended over $D$ as a $G$-bundle, there is a natural extension as a $\mathbb{T}_G$-bundle.

Therefore we have an alternative description of a gauge theory with a divisor defect: it is a theory based on a $G$-bundle $\mathcal{E}$ whose structure group is reduced to $\mathbb{T}_G$ along a divisor $D$. Note that this implies that in the Feynman path integral we must divide by gauge transformations which are $\mathbb{T}_G$ valued over $D$.

Just as in the case of surface operator, the defect we have described is not the most general but corresponds to the case where the parameters $\alpha$ are generic in $\mathbb{T}_G$. The more general case is parametrized by the pair $(\alpha , L)$, where $L$ is a Levi subgroup, defined as a subgroup of $G$ whose elements commute with $\alpha$. Clearly any Levi subgroup contains $\mathbb{T}_G$, which is indeed a minimal Levi subgroup. When a divisor defect is parametrized by $(\alpha , L)$, the only allowed gauge transformations are those which take values in $L$ when restricted to the divisor $D$.

In the following we will use an equivalent description using the correspondence between Levi subgroups of $G$ and parabolic subgroups of $G_{\mathbb{C}}$. Given the parameter $\alpha$ one can define the parabolic sub-algebra $\mathfrak{p}$ of $\mathfrak{g}_{\mathbb{C}}$ as the sub-algebra spanned by elements $x$ which obey
\begin{equation}
[ \alpha , x] = \ii \lambda \, x \ , \qquad \ \text{with} \ \lambda \ge 0 \ .
\end{equation}
The associated group $P \subset G_{\mathbb{C}}$ is called a parabolic subgroup; specifying a parabolic subgroup $P$ is equivalent to specifying the data $(\alpha, L)$. For example when $L=\mathbb{T}_G$, the corresponding parabolic group is a so called Borel subgroup and consists of upper triangular matrices of appropriate rank.  An equivalent definition of a parabolic subgroup of $G$ is as stabilizers of flags in $\mathbb{C}^n$. Recall that a flag is a sequence of subspaces
\begin{equation}
0 \subset U_1 \subset U_2 \subset \cdots \subset U_n = \mathbb{C}^r \ .
\end{equation}
The action of $G$ on the flag is given by
\begin{equation}
g \ \left( U_1, \, \dots , U_n \right) = \left( g \, U_1 , \, \dots , g \, U_n \right) \ .
\end{equation}
A \textit{complete} flag is characterized by $n=r$ and $\dim U_i = i$. In particular complete flags are stabilized by Borel subgroups. For example the standard complete flag is defined by the choice
\begin{equation}
U_i = \mathbb{C} \, e_1 \oplus \mathbb{C} \, e_2 \oplus \cdots \oplus \mathbb{C} \, e_i \ ,
\end{equation}
and it corresponds to the span of the first $i$ elements of the natural basis of $\mathbb{C}^r$. 

Therefore we can regard the gauge theory defined in the presence of a divisor defect, either as consisting of a $G$-bundle whose structure group is reduced to a Levi subgroup $L$ along the divisor $D$, or as consisting of an holomorphic $G_{\mathbb{C}}$-bundle whose structure group is reduced to a parabolic subgroup $P$ along $D$ (and of course similar definition hold for all the other fields). The latter perspective will be more useful to discuss BPS invariants.

\section{Donaldson-Thomas theory with divisor defects} \label{D-Tdivisor}

We will define Donaldson-Thomas theory with divisor defects as the study of the intersection theory of the moduli spaces of sheaves with a prescribed behavior along the divisor. Recall that in ordinary Donaldson-Thomas theory we are interested in solutions of the Donaldson-Uhlenbeck-Yau equations in (\ref{inste}), or more generally in stable torsion free coherent sheaves. 

As we have discussed we can think of gauge theory in the presence of a divisor defect as the theory of $G$-bundles on $X$ with structure group reduced to the Levi subgroup $L$ along $D$. Therefore is natural to consider the moduli problem associated with the equations
\begin{eqnarray} \label{instdiv} 
(F_A   -  2 \pi \alpha \ \delta_D)^{(0,2)} &=&
  \overline{\partial}\,_A^{\dagger} \rho
  \ , \nonumber\\[4pt]
(F_A   -  2 \pi \alpha \ \delta_D)^{(1,1)} \wedge t \wedge t + \big[\rho\,,\, \overline{\rho}\,\big] &=&
l~t \wedge t \wedge t \ 
\ ,
\end{eqnarray}
where once again we are only interested in solutions with $\rho = 0$ (and for simplicity we set $l=0$). Note that the source $\delta_D$ forces the gauge field to obey the required boundary conditions along $D$. Therefore we consider the moduli space
\begin{equation} \label{Mbundle2}
\mathcal{M}^{(\alpha)} \left( L ; X \right) = \left. \left\{ A \in  \mathcal{A} (X)  \, \Big{\vert} \ \begin{matrix}  
(F_A   -  2 \pi \alpha \, \delta_D)^{(0,2)} = 0 , \\[4pt]
(F_A   -  2 \pi \alpha \, \delta_D)^{(1,1)} \wedge t \wedge t = 0 \end{matrix} \
\right\} \right/ \mathcal{G}_D \ .
\end{equation}
As we have explained we only consider the group $\mathcal{G}_D$ of gauge transformations valued in $L$ along $D$. Here $\mathcal{A} (X)$ is the space of connections, which is an affine space modeled on $ \Omega^{0,1} \left( X ;  \mathrm{ad} \,  \mathcal{E}  \right)$. We will sometimes use the notation $\mathcal{M}^{(\alpha)}_{n, \beta, u ; r}$, where $(n , -\beta, u)  = (\mathrm{ch}_3 (\mathcal{E}) , \mathrm{ch}_2 (\mathcal{E}), c_1(\mathcal{E}))$, when we want to stress the topological numbers of $\mathcal{E}$.

The gauge theory perspective provides a natural integrand over this moduli space: the Euler class of the normal bundle $\mathrm{eul} (\mathcal{N}^{(\alpha)}_{n ,\beta, u ; r})$ which arises when restricting the instanton deformation complex (\ref{defcomplex}) to field configurations which obey (\ref{instdiv}). The gauge theory partition function has the form of a generating function of BPS invariants
\begin{align} \label{ZdtDiv}
\mathcal{Z}_{\mathrm{gauge}}^{(X , D)} (q, Q ; r) = & \sum_{n , \, \beta, \, u} \sum_{\frak{m} , \frak{h} , \frak{o}} \ q^k \ Q^{\beta} \ v^u \ \e^{ 2 \pi \ii \left( \eta^i \, m_i  + t^{D}_a \, \gamma^i \, o_i ^a + \sigma^i \, n_i \right)}
  \cr & \ \times  \int_{\mathcal{M}^{(\alpha)}_{n, \beta, u ; r} \left( L ; X \vert \{ \frak{m} , \frak{h} , \frak{o}  \} \right)} \ \mathrm{eul} (\mathcal{N}^{(\alpha)}_{n ,\beta, u ; r}) \ .
\end{align}
For completeness we have written down a slightly more general partition function, which include a set of ``theta-angles'' which further characterize the divisor defect. We will not discuss their origin in this survey, but refer the reader to \cite{Cirafici:2013nja}. These extra parameters are associated with the geometry of $D$: for example if $\mathbb{T}_G \simeq U(1)^r$, each rank one factor gives rise to a line bundle $\mathcal{L}$ on $D$, and for each factor $\mathcal{L}$ the above phase is
\begin{equation}
\exp 2 \pi \ii \left( \eta \int_{D} \mathrm{ch}_2 (\mathcal{L}) + \gamma \int_{D} c_1 (\mathcal{L}) \wedge k + \sigma \int_{D \cap D} c_1 (\mathcal{L}) \right) 
\end{equation}

It is however useful to adopt a different perspective. In the case without defects, instead of looking directly at the Donaldson-Uhlenbeck-Yau equations, one typically studies holomorphic bundles, trading the second equation for a stability condition. Furthermore relaxing the bundle condition leads to the moduli space of stable torsion free coherent sheaves. In line with these ideas we will propose that an alternative way of looking at solutions of (\ref{instdiv}) is by studying the moduli space of parabolic sheaves. This moduli space arises naturally when thinking of the relevant configurations in the presence of a defect as holomorphic bundles whose structure group is reduced to a parabolic group along $D$. Ideally one could hope that the moduli space of parabolic sheaves is a better behaved version (and perhaps a compactification) of (\ref{Mbundle2}). Regrettable these conjectures have not beed studied in the literature. We will therefore take a more practical approach and define directly Donaldson-Thomas type invariants associated with these moduli spaces; a more detailed discussion is in \cite{Cirafici:2013nja}.

A torsion free parabolic sheaf $\mathcal{E}$ is a torsion free sheaf on $X$ with a parabolic structure on $D$. The latter is the filtration
\begin{equation}
\mathcal{F}_\bullet \ : \ \ \mathcal{E} = \mathcal{F}_1 (\mathcal{E}) \supset \mathcal{F}_2 (\mathcal{E}) \supset \cdots \supset \mathcal{F}_l (\mathcal{E}) \supset \mathcal{F}_{l+1} = \mathcal{E} (-D) \ ,
\end{equation}
together with the ordered set of weights $0 \le a_1 < a_2 < \cdots < a_l \le 1$, which coincide with the parameters $\alpha$ up to a conventional normalization. Specifying the parabolic structure via a filtration is akin to the description of a parabolic group as the stabilizer of a certain flag. We will denote the moduli space of parabolic sheaves with $\mathcal{P}_{n , \beta , u ; r}^{ (\alpha)} (X , D \vert \{ \mathrm{ch} (\mathcal{F}_i (\mathcal{E})) \} ) $, or $\mathcal{P}_{n , \beta , u ; r}^{ (\alpha)}$ for simplicity; note that $\mathcal{F}_1 (\mathcal{E}) = \mathcal{E}$ and by definition $( \mathrm{ch}_3 (\mathcal{E}) , \mathrm{ch}_2 (\mathcal{E})  , c_1 (\mathcal{E}) = n , -\beta , u)$. A proper definition of these moduli spaces would require a notion of parabolic stability to select physical configurations; this can be done but we refer the reader to \cite{Cirafici:2013nja} for more details. All the parabolic sheaves considered in this note are assumed to be stable. We define Donaldson-Thomas invariants in the presence of a divisor defect as 
\begin{equation}
{\tt DT}^{(\alpha)}_{n,\beta, u;r} (X , D  ) = \int_{\mathcal{P}_{\beta , n, u ; r}^{(\alpha)}} \ \mathrm{eul} (\mathcal{N}^{(\alpha)}_{n ,\beta, u ; r}) \ , 
\end{equation}
and similarly for the gauge theory partition function (\ref{ZdtDiv}). In the following we will see that these definitions can be made very concrete in certain cases, and the enumerative invariants computed explicitly.

\section{Parabolic sheaves and orbifold sheaves} \label{parabolic}

We will now consider the simplest possible case, of a divisor defect on $\mathbb{C}^3$. For technical reasons it is easier to compactify $\mathbb{C}^3$ to $\mathbb{P}^1 \times \mathbb{P}^1 \times \mathbb{P}^1$ and define the defect on the divisor $\mathcal{D} = \mathbb{P}^1_{z_1} \times \mathbb{P}^1_{z_2} \times 0_{z_3}$ while denoting with $\mathcal{D}_{\infty} = \mathbb{P}^1_{z_1} \times \mathbb{P}^1_{z_2} \times \infty_{z_3} \sqcup \, \mathbb{P}^1_{z_1} \times \infty_{z_2} \times \mathbb{P}^1_{z_3} \sqcup \, \infty_{z_1} \times \mathbb{P}^1_{z_2} \times \mathbb{P}^1_{z_3}$ the divisor at infinity. We will discuss moduli spaces of objects on $\mathbb{P}^1 \times \mathbb{P}^1 \times \mathbb{P}^1$ which are trivialized on the divisor at infinity $\mathcal{D}_\infty$, corresponding to gauge fields on $\mathbb{C}^3$ which are flat at infinity.

To define the moduli space of parabolic sheaves we fix an $r$-tuple of integers $\mathbf{d} = (d_0 , \cdots , d_{r-1})$ which we will identify with the instanton numbers. A parabolic sheaf $\mathcal{F}_\bullet$ is defined by the flag of rank $r$ torsion free sheaves
\begin{equation}
\mathcal{F}_{0} (- \mathcal{D}) \subset \mathcal{F}_{-r+1} \subset \cdots \subset \mathcal{F}_{-1} \subset \mathcal{F}_0  \ ,
\end{equation}
such that $c_2 (\mathcal{F}_i)=0$ and $c_3 (\mathcal{F}_i) = - d_i$. The first Chern classes are fixed by the framing condition, that the sheaves are locally free at infinity:
\begin{equation} \label{framing}
\xymatrix{
 \mathcal{F}_{0} (- \mathcal{D}) \vert_{\mathcal{D}_{\infty}} \ar[r] \ar[d]^\simeq & \mathcal{F}_{-r+1} \vert_{\mathcal{D}_{\infty}}  \ar[r] \ar[d]^\simeq & \cdots \ar[r] \ar[d]^\simeq &   \mathcal{F}_0 \vert_{\mathcal{D}_{\infty}}  \ar[d]^\simeq \\
\mathcal{O}^{\oplus r}_{\mathcal{D}_{\infty}} (-\mathcal{D}) \ar[r] & W^{(1)} \otimes \mathcal{O}_{\mathcal{D}_{\infty}} \oplus \mathcal{O}^{\oplus r-1}_{\mathcal{D}_{\infty}} (-\mathcal{D}) \ar[r] & \cdots \ar[r]  & W^{(r)} \otimes \mathcal{O}_{\mathcal{D}_{\infty}}
}
\end{equation}
This condition simply means that the parabolic structure and the boundary conditions at infinity should be compatible since the divisor at infinity $\mathcal{D}_{\infty}$ and $\mathcal{D}$ have a non trivial intersection. Note that at infinity $\mathcal{F}_0$ is isomorphic to the rank $r$ locally free sheaf $\mathcal{O}^{\oplus r}$. By picking a basis we identify $\mathcal{O}^{\oplus r}$ with the vector space $W^{r} = \langle w_1 , \dots , w_r \rangle$. Similarly $W^{(i)} = \langle w_1 , \dots , w_i \rangle$ for $i = 1 , \dots , r$ are $i$-dimensional vector spaces. In this notation the parabolic group stabilizes the flag
\begin{equation} \label{flag}
W^{(1)} \subset W^{(2)} \subset \cdots \subset W^{(r)} = W \ .
\end{equation}
Note that in the case of a Borel subgroup this would be the standard complete flag. For simplicity we will assume that this is the case in the remaining of this note. The most general case is discussed in \cite{Cirafici:2013nja}, where it is shown how equation (\ref{framing}) can be modified to allow for more general flags, and how the correspondence with orbifold sheaves, to be discussed momentarily, can be generalized accordingly. The construction outlined above is an higher dimensional  generalization of the construction of \cite{feigin,finkel,negut}.

The moduli space $\mathcal{P}_{\mathbf{d}}$ can be constructed very explicitly as the fixed component of $\mathcal{M}^{\rm inst}_{n , 0 ; r} (\mathbb{C}^3)$ under the action of a discrete group $\Gamma$. This is an higher dimensional generalization of the correspondence between parabolic sheaves and orbifold sheaves, or sheaves which are $\Gamma$-equivariant, used in \cite{feigin,finkel,negut}. This correspondence will allow us to use the formalism introduced in Section \ref{instantons} to study the new enumerative invariants. We will now sketch this correspondence.

Let the group $\Gamma = \mathbb{Z}_r$ act on $\mathbb{C}^3$ as
\begin{equation} \label{gammadiv}
(z_1 , \, z_2 , \, z_3 ) \longrightarrow (z_1 , \, z_2 , \, \omega \, z_3) \ ,
\end{equation}
where $r \in \mathbb{Z}$ and $\omega = \e^{\frac{2 \pi \ii}{r}}$. This action is chosen in such a way that $\mathcal{D}$ is invariant. Construct an analog of the flag (\ref{flag}) we let $\Gamma$ act also on $W = \langle w_1 , \dots , w_r \rangle$ as $\gamma (w_l) = \e^{\frac{2 \pi \ii l}{r}} \, w_l$. The isotypical decomposition of $W$ is obtained by summing over all the irreducible representations of $\hat{\Gamma}$
\begin{equation}
W = \bigoplus_{a \in \hat{\Gamma}} W_{a} \otimes \rho^{\vee}_a \ .
\end{equation}
Since we are limiting ourselves to the case where the flag stabilized is the standard complete flag, or equivalently the parabolic group is a Borel group, each factor has $\dim W_{a} = 1$. The identification between the isotypical decomposition under $\Gamma$ and the flag (\ref{flag}) is given by
\begin{equation} \label{Wdec2}
W^{(i)} = \bigoplus_{a=0}^{i-1} W_a  \ .
\end{equation}
The framing condition can be now equivalently expressed in terms of the vector spaces $W_a$, $a=0, \dots , r-1$.
Define the covering map 
\begin{eqnarray}
\sigma : \mathbb{P}^1_{z_1} \times \mathbb{P}^1_{z_2} \times \mathbb{P}^1_{z_3} &\longrightarrow&  \mathbb{P}^1_{z_1} \times \mathbb{P}^1_{z_2} \times \mathbb{P}^1_{z_3} \cr
(z_1 , \, z_2 , \, z_3 ) &\longrightarrow& (z_1 , \, z_2 , \, z_3^r ) \ .
\end{eqnarray}
As in \cite{feigin,finkel} this map gives an isomorphism which identifies a parabolic sheaf $\mathcal{F}_\bullet$ in $\mathcal{P}_{\mathbb{d}}$ with a $\Gamma$-equivariant sheaf $\tilde{\mathcal{F}}$ in $\mathcal{M}^{\rm inst}_{\mathbf{d} , 0 ; r} (\mathbb{C}^3)^{\Gamma}$. Roughly speaking the dictionary is given by
\begin{equation} \label{flagfromGamma}
\mathcal{F}_{k} = \sigma_* \left( \tilde{\mathcal{F}} \otimes \mathcal{O}_X (k \, \mathcal{D}) \right)^{\Gamma} \ .
\end{equation}
see \cite{Cirafici:2013nja} for a more detailed discussion. In plain words given a $\Gamma$-equivariant sheaf $\tilde{\mathcal{F}}$, we construct a flag from the isotypical components; viceversa given a parabolic sheaf $\mathcal{F}_\bullet$ we construct a $\Gamma$-equivariant sheaf by pulling back each element of the flag via $\sigma$ and interpreting the result as an element of an isotypical decomposition. The correspondence is rather simple, although one has to be rather careful in the precise details in order to obtain the correct Chern classes. The precise identification between the moduli spaces is
\begin{equation}
\mathcal{M}^{\rm inst}_{n , 0 ; r} (\mathbb{C}^3)^{\Gamma} = \bigcup_{|\mathbf{d}| = n} \ \mathcal{P}_{\mathbf{d}} \ .
\end{equation}

\section{Quivers and divisor defects} \label{quiverdivisor}

We have reduced the case of studying divisor defects on $\mathbb{C}^3$ to an equivariant problem. Therefore we can simply apply verbatim the formalism we have discussed in Section \ref{instantons} with the orbifold action $\Gamma$ given by (\ref{gammadiv}). We stress that this is just a property of the formalism and no orbifold singularity is present in the physical theory.

The relevant quiver quantum mechanics model arises from the ADHM formalism by decomposing the ADHM data according to the action of $\Gamma$. The vector spaces $V$ and $W$ decompose as
\begin{equation} \label{VWdiv}
V = \bigoplus_{a\in\widehat{\Gamma}}\, V_a \otimes
\rho_a^{\vee} \ , \qquad W = \bigoplus_{a\in\widehat{\Gamma}}\, W_a \otimes \rho_a^{\vee} \ ,
\end{equation}
Recall that we are considering the simplest divisor defect, corresponding to a Borel group; the generalization to an arbitrary parabolic group simply amounts in defining the vector spaces $W_a$ in such a way that the appropriate flag is recovered using the dictionary (\ref{Wdec2}), as discussed in greater details in \cite{Cirafici:2013nja}. The bosonic field content of the quiver quantum mechanics is again
\begin{eqnarray}
(B_1 , B_2 , B_3 , \varphi) \in\mathrm{Hom}_{\Gamma} (V,V) \qquad \mbox{and} \qquad
I\in\mathrm{Hom}_{\Gamma} (W,V) \ .
\end{eqnarray}
In this case the only non trivial maps are
\begin{eqnarray}
B_{1,2}^{a} \, &:& \, V_{a} \ \longrightarrow \ V_{a} \ , \nonumber \\[4pt]
B_{3}^{a} \, &:& \, V_{a} \ \longrightarrow \ V_{a+1} \ , \nonumber \\[4pt]
\varphi^{a} \, & : & \, V_{a} \ \longrightarrow \ V_{a+1} \ , \nonumber \\[4pt]
I^{a} \, & : & \, W_{a} \ \longrightarrow \ V_{a} \ .
\end{eqnarray}
The BRST transformations read
\begin{equation}
\begin{array}{rllrl}
\mathcal{Q} \, B_\alpha^{a} &=& \psi_\alpha^r & \quad \mbox{and} \qquad \mathcal{Q} \, \psi_\alpha^{a}& = \ [\phi , B_\alpha^{a}] -
\epsilon_\alpha\,
B_\alpha^{a} \ , \\[4pt] \mathcal{Q} \, \varphi^{r}& =& \xi^{r} & \quad \mbox{and} \qquad
\mathcal{Q} \, \xi^{a} &= \ [\phi ,\varphi^{a}] - (\epsilon_1 + \epsilon_2 + \epsilon_3 ) \varphi^{a}  \ , \\[4pt]
\mathcal{Q} \, I^{a} &=& \varrho^{a} & \quad \mbox{and} \qquad \mathcal{Q} \, \varrho^{a} & = \
\phi \,I^{a} - I^{a} \,\mathbf{a}^{a} \ ,
\end{array}
\end{equation}
where $\mathbf{a}^{a}$ is a vector whose components are the Higgs field eigenvalues $a_l$; these are associated via (\ref{VWdiv}) with the irreducible representation $\rho_a$. Note that we do not impose any condition on 
$\epsilon_1 + \epsilon_2 + \epsilon_3 $ since $\Gamma$ is not a subgroup of $SL (3 , \mathbb{C}^3)$. These data correspond to the quiver
\begin{equation} \label{ADHMgen}
\vspace{4pt}
\xymatrix@C=15mm{
 \cdots  \ar@//[r]^{B^{a-2}_3}
\ar@{.>}@/_/[r]_{\varphi^{a-2}} 
& \ V_{a-1 }\ \bullet \ \ar@(u,ul)_{B^{a-1}_2} \ar@(u,ur)^{B^{a-1}_1} \ar@//[r]^{B^{a-1}_3}
\ar@{.>}@/_/[r]_{\varphi^{a-1}} 
& \ V_a \ \bullet \ \ar@(u,ul)_{B^a_2} \ar@(u,ur)^{B^a_1} \ar@//[r]^{B^a_3}
\ar@{.>}@/_/[r]_{\varphi^a} 
& \ V_{a+1} \ \bullet \ \ar@(u,ul)_{B^{a+1}_2} \ar@(u,ur)^{B^{a+1}_1} \ar@//[r]^{B^{a+1}_3}
\ar@{.>}@/_/[r]_{\varphi^{a+1}} & \cdots
 \\ \\
&  W_{a-1} \ \bullet \ar@//[uu]^{I^{a-1}} 
& W_a \ \bullet \ar@//[uu]^{I^a} 
& W_{a+1} \ \bullet \ar@//[uu]^{I^{a+1}} 
&
}
\vspace{4pt}
\end{equation}
The moduli space of BPS configurations in the presence of the defect is now the moduli stack
\begin{equation}
\mathcal{M}_{\Gamma} (\mathbf{n} ,\mathbf{r}) = \Big[ Rep_{\Gamma} (\mathbf{n} ,\mathbf{r} ; B) \,
\Big/ \, \prod_{a\in\widehat \Gamma}\, GL(n_a,\mathbb{C}) \Big] \ ,
\end{equation}
where $Rep_{\Gamma} (\mathbf{n} ,\mathbf{r} ; B)$ is the sub variety of the representation space
\begin{equation}
{Rep}_{\Gamma}  (\mathbf{n} ,\mathbf{r})  =  \mathrm{Hom}_{\Gamma} (V , Q \otimes V) \ 
 \oplus \ \mathrm{Hom}_{\Gamma} (V , \mbox{$\bigwedge^3$} Q \otimes V) \ 
\oplus \ \mathrm{Hom}_{\Gamma} (W , V) \ ,
\end{equation}
cut out by the equations
\begin{equation} \label{surfADHM}
\begin{array}{rllrl}
B_1^a \, B_2^a & - & B_2^a \, B_1^a&=& 0 \, ,\\
 B_1^{a+1} \, B_3^a & - & B_3^a \, B_1^a &=& 0 \, , \\
 B_2^{a+1} \, B_3^a & - & B^a_3 \, B_2^a &=& 0 \, .
\end{array}
\end{equation}
The form of the regular representation $Q = \rho_0 \oplus \rho_0 \oplus \rho_1$ follows from the definition of the $\Gamma$-action in (\ref{gammadiv}). Now the enumerative invariants can be defined and computed via virtual localization as
\begin{align} \label{DTdivFinal}
{\tt DT}_{n , r}^{\mathcal{D}} \left( \mathbb{C}^3 \, \vert \, \epsilon_1 , \epsilon_2 , \epsilon_3 , a_l \right) = & \int_{ [ \mathcal{M}_{\Gamma} (\mathbf{n} ,\mathbf{r}) ]^{\mathrm{vir}} } \ 1 
\\ \nonumber = & \sum_{[\mathcal{E}_{\vec{\pi}}] \in \mathcal{M}_{\Gamma} (\mathbf{n} ,\mathbf{r})^{\mathbb{T}^3 \times U(1)^r} } \ \frac{1}{\mathrm{eul}_{\mathbb{T}^3 \times U(1)^r } \left( T_{[\mathcal{E}_{\vec{\pi}}]}^{\mathrm{vir}}  \mathcal{M}_{\Gamma} (\mathbf{n} ,\mathbf{r})   \right) }  \ .
\end{align}
where the fixed point set of the toric action $\mathbb{T}^3 \times U(1)^r $ is given by vectors of plane partitions which carry a $\Gamma$-action. The only difference from the orbifold case is that the new invariants depend explicitly on the parameters in the Cartan sub algebra of $\mathbb{T}^3 \times U(1)^r$. In other words in the case of $\mathbb{C}^3$ we are defining an equivariant version of Donaldson-Thomas theory. The right hand side of (\ref{DTdivFinal}) can be computed explicitly from the character at a fixed point
\begin{align}
& \mathrm{ch}_{\mathbb{T}^3 \times U(1)^r} \left(  T_{\vec{\pi}}^{\mathrm{vir}}   \mathcal{M}_{\Gamma}  (\mathbf{n} ,\mathbf{r}) \right)
\\ \nonumber & = \left( W_{\vec\pi}^\vee \otimes V_{\vec\pi} -
\frac{ {V}_{\vec\pi}^\vee \otimes W_{\vec\pi} }{t_1 \, t_2 \, t_3} + \frac{ (1-t_1)\, (1-t_2)\,
(1-t_3) }{t_1 \, t_2 \, t_3} ~ {V}^\vee_{\vec\pi} \otimes V_{\vec\pi} \right)^{\Gamma} \ .
\end{align}
We refer the reader to \cite{Cirafici:2013nja} for explicit formulae.

\section{Higher dimensional theories} \label{higher}

The ideas we have discussed so far have a broad range of applications. In particular one could take a generic quantum field theory, topological or not, introduce a class of defects and investigate if they are associated with new BPS invariants. For example, BPS states in supersymmetric field theories are typically associated with certain moduli spaces of field configurations. Since defects can be thought of as particular boundary conditions, it can happen that when the theory is modified by the presence of the defect, the moduli spaces of BPS configurations are modified as well. The intersection theory of these new moduli spaces is an interesting mathematical problem and can in principle provide new BPS invariants.

Typically one begins with a certain manifold $M_d$ of real dimension $d$, with certain structures related to its holonomy, or its reduction. The relevant instanton equations have the form
\begin{equation}
\lambda \, F_{\mu \nu} = \frac12 T_{\mu \nu \rho \sigma} \, F^{\rho \sigma}
\end{equation}
for a field strength $F$, where $\lambda$ is a constant and the antisymmetric tensor $T_{\mu \nu \rho \sigma}$ is responsible for the reduction of the holonomy of $M_d$. To be more concrete, we set $d=8$ and pick as tensor $T$ the holomorphic $(4,0)$ form $\Omega$ so that the holonomy is reduced to $SU(4)$. 

We use $\Omega$ to define the operator $*$ on $M_8$ as
\begin{equation}
* \ : \ \Omega^{0,p} (M_8)  \longrightarrow \Omega^{0,4-p} (M_8) \ ,
\end{equation}
via the pairing
\begin{equation}
\langle \alpha , \beta \rangle = \int_{M_8} \ \Omega \wedge \alpha \wedge * \beta \ .
\end{equation}
Let $\Omega^{0,2}_{\pm} (M_8)$ be the eigenspaces of $*$ and $P_{\pm}$ be the natural projection. Given an holomorphic bundle $\mathcal{E}$, we say that a connection $\overline{\partial}_A$ with $F_A^{0,2} = \overline{\partial}_A^2$ is  \textit{holomorphic anti-self-dual} if $P_+ F_A^{0,2}=0$. With these data we construct the following elliptic complex of adjoint valued differential forms
\begin{equation}
\label{8dcomplex}
\xymatrix@1{0 \ar[r] & \Omega^{0,0} ( M_8 , \mathrm{ad}\,  \mathcal{E})
\ar[r]^{\hspace{-0.2cm} \overline{\partial}_A} & ~\Omega^{0,1} ( M_8 , \mathrm{ad}\, \mathcal{E}) 
\ar[r]^{\hspace{0.2cm} P_+ \overline{\partial}_A} & \Omega^{0,2} ( M_8 , \mathrm{ad}\, \mathcal{E}) \ar[r] & 0 } \ .
\end{equation}
One can construct a cohomological theory by gauge fixing the topological invariant \cite{Baulieu:1997jx}
\begin{equation}
S_8 = \int_{M_8} \Omega \wedge \mathrm{Tr} (F_A^{0,2} \wedge F^{0,2}_A) \ .
\end{equation}
This theory localizes onto the moduli space of holomorphic anti-self-dual connections in the topological sector with $S_8$ fixed. The intersection theory of this moduli space has not been rigorously defined, yet this theory is believed to give a higher dimensional generalization of Donaldson theory \cite{Baulieu:1997jx}.

We can now apply the formalism we have introduced almost verbatim, to define a divisor defect on a co-dimension two divisor $D_6$, and study the associated modified moduli space
\begin{equation} 
\mathcal{M}^{\alpha} (L ; M_8) = \left. \left\{ A \in  \Omega^{(0,1)} \left( M_8 ; \mathrm{ad} \,  \mathcal{E}  \right)  \, \Big{\vert} \   
P_+ \, (F_A   -  2 \pi \alpha \delta_{D_6})^{(0,2)} = 0 , \text{stable}
\right\} \right/ \mathcal{G}_{D_6} \ ,
\end{equation}
which parametrizes holomorphic anti-self-dual connections whose structure group is reduced to a Levi group $L$ along $D_6$, modulo gauge transformations which take values in $L$ on $D_6$, with an appropriate stability condition. Similarly we can introduce the moduli spaces $\mathcal{P}^{\alpha} (M_8 ; D_6)$ of stable torsion free sheaves with a parabolic structure over $D_6$. These moduli spaces have not so far been studied due to challenging technical difficulties; we believe that some progress in this direction can be made using quantum field theory techniques, possibly by defining new BPS invariants. A proposal to solve many difficulties concerning Donaldson-Thomas theory on Calabi-Yau 4-fold was recently put forward in \cite{leung}. It is possible that a modification of the construction of \cite{leung} could lead to a proper definition of defects in higher dimensional theories.

\section{Line defects in $\mathcal{N}=2$ $4d$ QFT} \label{line}

We shall now discuss another example: BPS invariants which arise in the case of a four dimensional quantum field theory with $\mathcal{N}=2$, where we modify the theory by introducing a line defect \cite{Gaiotto:2010be,Chuang:2013wt,Cordova:2013bza,Xie:2013lca,Cirafici:2013bha,CDZ,CDZ2}. These theories have moduli spaces of vacua, and we will consider a generic point $u$ in the Coulomb branch $\mathcal{B}$ where the gauge symmetry is spontaneously broken down to its maximal torus. The Coulomb branch is divided into chambers and the counting of stable BPS states in each chamber differs by the application of the wall-crossing formula. We will denote by $\Gamma_g$ the lattice of electric and magnetic charges. The lattice of charges is endowed with an antisymmetric integral pairing $\langle  \ , \ \rangle : \Gamma_g \times \Gamma_g \longrightarrow \mathbb{Z}$.

We think of a line defect as a point defect located at the origin of $\mathbb{R}^3$ which extends in the time direction. Physically it can be modeled on a non-dynamical heavy particle of charge $\gamma_f$. This defect can be chosen to be supersymmetric. The Hilbert space of the theory is modified by the presence of the defect. Roughly speaking it can be identified with the cohomology of a moduli space of BPS configuration in the presence of the line defect
\begin{equation}
\mathcal{H}_{L,u} = \bigoplus_{\gamma \in \Gamma} \mathcal{H}_{L,\gamma,u} = \bigoplus_{\gamma \in \Gamma} \bigoplus_{p,q} \, H^{p,q} (\mathcal{M}^{\tt BPS} (L , \gamma , u)) \, .
\end{equation}
Here we have used the fact that Hilbert spaces in quantum field theories are naturally graded by the electromagnetic charge $\gamma$ as measured at infinity. 

It turns out that the relevant moduli space of BPS configuration can be once again identified with the moduli space of stable framed quiver representations. This quiver is constructed by taking a positive basis of charges of $\Gamma_g$, say $\{ \gamma_i \}$ and associating each element with a node of the quiver \cite{Alim:2011kw}. The numbers of arrows from node $\gamma_i$ to node $\gamma_j$ is given by $\langle \gamma_i , \gamma_j \rangle$. The framing node is associated with the charge $\gamma_f$ of the defect, and connected with the rest of the quiver by $\langle \gamma_f , \gamma_i \rangle$ arrows for $i \in \mathsf{Q}_0$. Now one can define BPS invariants by integration over the moduli spaces of framed representations with a suitable stability condition. These integrals can be defined via virtual localization as we have done previously, with respect to a natural torus $\mathbb{T} $ which acts on the maps of a quiver representation by rescaling. If we denote the relevant moduli space of quiver representations as $\mathcal{M}_{\mathbf{d}} (\widehat{\mathsf{Q}})$, we have
\begin{equation}
{\tt DT}_{\mathbf{d}} = \int_{[\mathcal{M}_{\mathbf{d}} (\widehat{\mathsf{Q}})]^{vir}} \ 1 = \sum_{\pi \in \mathcal{M}_{\mathbf{d}} (\widehat{\mathsf{Q}})^{\mathbb{T}} } \ (-1)^{\mathrm{dim T_{\pi} \mathcal{M}_{\mathbf{d}} (\widehat{\mathsf{Q}})}} \, .
\end{equation}
In the above formula the dimension of a representation $\mathbf{d} = (d_1 , \cdots , d_k)$ is related to the charge of a BPS state by $\gamma = \gamma_f + \sum_{i \in \mathsf{Q}_0} d_i \gamma_i$. The fixed points $\pi$ can be given an explicit combinatorial classification.  A full treatment of this construction, as well as its relation with the theory of cluster algebras, is given in \cite{CDZ}.

\section{Discussion} \label{discussion}

In this note we have provided a brief survey on certain enumerative invariants of Donaldson-Thomas type which can be computed via framed quivers. The main theme is that in certain chambers, at large radius and on the noncommutative crepant resolution, these invariants can be understood as generalized instanton configurations in a topological quantum field theory. The instanton computation reduces to the study of an effective quantum mechanics which is based on a framed quiver. The problem becomes completely algebraic and is reduced to the study of the representation theory of the framed quiver.

We have argued that this construction is modified by the presence of defects in the topological field theory. We have discussed in some detail the case of a divisor defect, which corresponds geometrically to studying torsion free coherent sheaves with a parabolic structure along a divisor. In certain simple cases, the formalism based on quivers can be adapted and used efficiently. In the more general case, the existence of a consistent enumerative problem is still conjectural. 

More in general, one can advocate a broader program which consists in understanding how Donaldson-Thomas theory is modified by the presence of defects in the physical theories. We have given few examples of how this program can be carried on. We expect that this line of investigation will lead to interesting mathematical structures as well as physical applications.

\section*{Acknowledgments}

I thank M. Del Zotto,  A. Kashani-Poor, A. Sinkovics and R.J. Szabo for discussions and collaborations on these and related topics; I also thanks J. Manschot and T. Strobl for the invitation to contribute to this volume. My work is supported by the Funda\c{c}\~{a}o para a Ci\^{e}ncia e Tecnologia (FCT/Portugal) via the \textit{Investigador FCT} program (contract IF/01426/2014/CP1214/CT0001), the project PEst-OE/EEI/LA0009/2013 and grants PTDC/MAT/119689/2010 and EXCL/MAT-GEO/0222/2012.

\end{document}